%% file: manus.tex
\documentclass[a4paper,11pt]{article}

\usepackage{geometry}										
\usepackage[english]{babel}									
\usepackage[utf8]{inputenc}									
\usepackage[T1]{fontenc}									

\usepackage{lmodern}										

\usepackage{lineno}

\usepackage{amssymb,amsfonts,amsmath,amsthm}
\usepackage{booktabs}
\usepackage{hyphenat}
\usepackage{bbm}
\usepackage{multirow}
\usepackage{graphicx}

\usepackage{caption}
\usepackage{hyperref}
\usepackage{subfigure}
\usepackage{xcolor}
\usepackage{placeins}
\usepackage{mathtools}
\usepackage{adjustbox}
\usepackage{stfloats} 

\makeatletter
\let\oldabs\abs
\def\abs{\@ifstar{\oldabs}{\oldabs*}}
\let\oldnorm\norm
\def\norm{\@ifstar{\oldnorm}{\oldnorm*}}
\makeatother

\newcommand{\ve}[1]{\boldsymbol{#1}}			
\newcommand{\ds}{\displaystyle}

\newcommand{\pdf}{\textit{pdf}}
\newcommand{\ecs}{\mathcal{X}}

\usepackage{authblk}												

\title{UQ state-dependent framework for seismic fragility assessment of industrial components}
\author[1]{Chiara Nardin \thanks{chiara.nardin@unitn.it}}
\author[2]{Stefano Marelli}
\author[1]{Oreste S. Bursi}
\author[2]{Bruno Sudret}
\author[1]{Marco Broccardo}
\affil[1]{Department of Civil, Environmental and Mechanical Engineering, University of Trento, Italy}
\affil[2]{Chair of Risk, Safety and Uncertainty Quantification, ETH Z\"{u}rich, Switzerland}

\date{\today}

\usepackage{microtype}										

\usepackage{indentfirst}									

\usepackage{caption}
\usepackage{subcaption}

\usepackage{natbib}
\usepackage{hyperref}
\hypersetup{
pdftitle = {UQ state-dependent framework for seismic fragility assessment of industrial components},
pdfauthor = {Chiara Nardin, Stefano Marelli, Oreste S. Bursi, Bruno Sudret, Marco Broccardo},
pdfsubject = {},
pdfkeywords = {state-dependent fragility functions, surrogate modelling, polynomial chaos expansion, industrial components, uncertainty quantification, seismic vulnerability},
colorlinks = false,                                                                                                                                                                          
}

\usepackage[capitalize]{cleveref}								
\creflabelformat{equation}{#2(#1)#3}								
\crefrangelabelformat{equation}{#3(#1)#4 to #5(#2)#6}						
\labelcrefformat{subequation}{#2(#1)#3}								
\labelcrefrangeformat{subequation}{#3(#1)#4 to #5(#2)#6}					

\graphicspath{{./figures}} 

\begin{document}

\maketitle

\begin{abstract}
In this study, we propose a novel surrogate modelling approach to efficiently and accurately approximate the response of complex dynamical systems driven by time-varying Recently, there has been increased interest in assessing the seismic fragility of industrial plants and process equipment. This is reflected in the growing number of studies,  community-funded research projects and experimental campaigns on the matter.
Nonetheless, the complexity of the problem and its inherent modelling, coupled with a general scarcity of available data on process equipment, has limited the development of risk assessment methods. In fact, these limitations have led to the creation of simplified and quick-to-run models.
In this context, we propose an innovative framework for developing state-dependent fragility functions. This new methodology combines limited data with the power of metamodelling and statistical techniques, namely polynomial chaos expansions (PCE) and bootstrapping. Therefore, we validated the framework on a simplified and inexpensive-to-run MDoF system endowed with Bouc-Wen hysteresis. 
Then, we tested it on a real nonstructural industrial process component. Specifically, we applied the state-dependent fragility framework to a critical vertical tank of a multicomponent full-scale 3D steel braced frame (BF). The seismic performance of the BF endowed with process components was captured by means of shake table campaign within the European SPIF project. Finally, we derived state-dependent fragility functions based on the combination of PCE and bootstrap at a greatly reduced computational cost.
\end{abstract}

\newpage
\section{Introduction}\label{ch1:intro}
\subsection{Background and motivation}
Assessing structural and non-structural component vulnerability to earthquakes is a key step in modern probabilistic seismic risk assessment \cite{bib:Du2021}. The PEER Performance-based earthquake engineering (PBEE) framework has gained significant momentum in this field, thanks to its inherently versatile formulation. Its strength lies in a simple yet effective implementation of the total probability theorem, which allows one to decouple and then combine the output of probabilistic seismic hazard analysis (PSHA) with fragility, damage, and loss analysis.
To this end, the fragility analysis step offers the critical link between seismic hazard and structural modelling, since it estimates conditional probability of attaining or exceeding a specified damage state (DS), given an intensity measure (IM) of earthquake motion. Initially introduced for nuclear safety evaluation \cite{bib:richardson1980}, fragility curves are nowadays widely used, ranging from assessment of collapse risk \cite{bib:eads2013} to loss estimation \cite{bib:rossi2020}, from resilience quantification at individual scale \cite{bib:cimellaro2010} to community scale \cite{bib:burton2016}, etc. In recent years, several novel methodological contributions to fragility analysis have been made. They include the development of multivariate fragility functions \cite{bib:du2020}, the introduction of seismic fragility analysis based on a combination of real \cite{bib:galasso2020} and artificial ground motions and surrogate modelling \cite{bib:abbiati2021}, and the consideration of both state \cite{bib:iervolino2015} and time-dependent fragility \cite{bib:zio2020}, \cite{bib:padgett2010}.
Nonetheless, the majority of the past research focus was committed to characterise the damage transition from a pristine state, i.e., no seismic damage, to a more severe damage state for structures subjected to a single ground motion \cite{bib:du2020}, \cite{bib:galasso2020}, \cite{bib:padgett2010}.
Conversely, significantly less research has been devoted to state-dependent fragility modelling,
which (i) can capture the damage accumulation due to sequential seismic events, and (ii) enables fragility estimations of structures with different initial damage states. To the best of our knowledge, only the first attempt has been made by \cite{bib:iervolino2015}, \cite{bib:iervolino2017}, and \cite{bib:jia2018}.
As a matter of fact, the compounding effect of damage accumulation and disruption caused by sequences of earthquakes (such as in Wenchuan (2008), Tohoku (2011) and central Italy (2016)), has highlighted the importance of accurately capturing the effect of irreversible damage accumulation for a reliable risk assessment. 
In this respect recent research, see, among others,  \cite{bib:abdelnaby2018} and \cite{bib:kassem2019}, deeply investigated the effects of sequential seismic events and damage accumulation on the performance of RC frames, and computed the corresponding fragility curves. The results clearly demonstrated a substantial difference in seismic loss assessment. Similar conclusions are reached in~\cite{bib:zhang2018} for fixed and base-isolated steel frame structures.
In parallel, consistent procedures for selecting ground motions for event sequences have been the subject of thorough studies. Utilizing knowledge from prior work and literature, the authors in \cite{bib:zhang2018} provided recommendations for selecting record pairs for sequential response history analysis and seismic performance assessment. In particular, the authors recommended the use of a seismic sequence (SS) of main-shock (MS) and after-shock (AS) record pairs, because they naturally capture and preserve within-sequence correlations. Besides, the use of SS-MS-AS record sets is preferred since databases with MS-AS records continue to expand and develop. However, to adequately populate the event sequences, ground motion models may serve the purpose.
Advances in non-linear structural response simulations have yielded more accurate modelling of complex and multi-mode systems, allowing for gaining better insights into critical issues and performance behaviour of structures and installed non-structural components (NSCs). A more thorough discussion of these issues is reported in \cite{bib:quinci2023}. Recently, significant efforts have been made to further investigate the coupling effects between the main structure and NSCs, both numerically \cite{bib:filiatrault2001} and \cite{bib:debiasio2015}, and experimentally \cite{bib:nardin2022}, \cite{bib:mosqueda2009}. An important milestone can be found in \cite{bib:NIST2017} that summarizes a year-long study, which collected and documented the body of available knowledge related to the seismic performance of NSCs for civil and industrial structures. Once again, since NSCs for industrial plants account for the majority of direct property losses due to earthquake damage \cite{bib:filiatrault2001}, \cite{bib:NEHRP2015}, they were identified as a top priority in seismic risk assessment. However, to limit costly and disruptive non-structural damage is challenging, due to the need of predictive non-linear dynamic analysis for complex systems with strong coupling interactions with NSCs. They are mostly limited by available computational resources. 
In addition, the computation of fragility analysis requires a large number of non-linear time history analyses (NLTHA), limiting \textit{de facto} the total number of possible simulations. 
This issue is central in simulation-based uncertainty quantification (UQ), which is tackled by replacing the computationally expensive NLTHA of FE models with an equivalent surrogate model, as in \cite{bib:rocquigny2008}, \cite{bib:sudret2007}, \cite{bib:sudret2017}.
In the context of fragility assessment, researchers have adopted different metamodelling techniques to offset the computational burden related to the large number of simulations needed. For example, the authors in ~\cite{bib:Du2021} proposed a fragility modelling approach based on artificial neural networks for the initial and final damage classification. Moreover, \cite{bib:abbiati2021} adopted hierarchical Kriging to compute a multi-fidelity surrogate that fuses the predictions of multiple models for fragility assessment. 
Among the catalogue of families of surrogate models, polynomial chaos expansions (PCE) \cite{bib:blatman2008}, Bayesian networks \cite{bib:lu2021}, support vector machines \cite{bib:hurtado2007}, and artificial neural networks \cite{bib:chakraborty2023} have arguably become the most popular in civil engineering, since they provide more than just an approximation to the underlying computational model. 
In particular, they additionally yield analytical estimates of the response moments of the model, sensitivity indices or confidence levels for their own predictions. For example, \cite{bib:zhu2023} estimated the full conditional probability distribution of EDP conditioned on IMs by means of stochastic PCE.
An in-depth literature review of the current state-of-the-art in surrogate modelling for reliability assessment is presented in \cite{bib:texeira2021} and \cite{bib:moustapha2022}.
Undoubtedly, one of the most significant benefits of surrogate models is their computational efficiency after training, allowing for millions of model evaluations per second, even on common end-user hardware. This enables the estimation of empirical fragility functions, as addressed in the literature by the global earthquake model (GEM) \cite{bib:gem2015} and \cite{bib:ioannou2012}.

\subsection{Scope and core contribution}
On these premises, this paper presents a novel UQ-based framework for efficiently deriving state-dependent fragility curves of industrial system components. This approach blends data from a reduced number of complex and expensive sequential NLTHAs with cutting-edge metamodelling techniques. 
In detail, the paper is organised as follows. 
In Section~\ref{ch2:methodology}, we present the heuristics and the novel framework for state-dependent fragility assessment. In Section~\ref{ch3:benchmark}, we validate the methodology on an inexpensive-to-evaluate benchmark case study. Specifically, state-dependent fragility functions derived from brute-force Monte Carlo Simulation (MCS) are compared with surrogate-based MCS. In Section~\ref{ch4:application}, we applied the framework to a real industrial case study. Precisely, the case study deals with a critical non-structural component (NSC) installed on the steel braced frame (BF) substructure, i.e. a vertical tank, as part of the project SPIF (\cite{bib:nardin2022}). Finally, we provide in Section~\ref{ch5:conclusion} a summary of the main findings as well as future development perspectives.

\section{A state-dependent fragility analysis framework}\label{ch2:methodology}
Limited by the availability of time and information, risk-informed assessments are commonly carried out on the basis of simplified and quick-to-evaluate models. However, as highlighted in the recent \cite{bib:NIST2017} report devoted to industrial facilities, these proven strategies lead to neither economically viable nor rational designs.
On these premises, this Section presents the heuristic of a novel framework for system vulnerability assessment. Specifically, based on a combination of experimental data and surrogate models, the proposed methodology enables the computation of state-dependent fragility curves that consider several aspects of the problem. 

\begin{figure}[!ht]
	\centering
	\begin{subfigure}
		\centering
		\includegraphics[width=0.90\columnwidth]{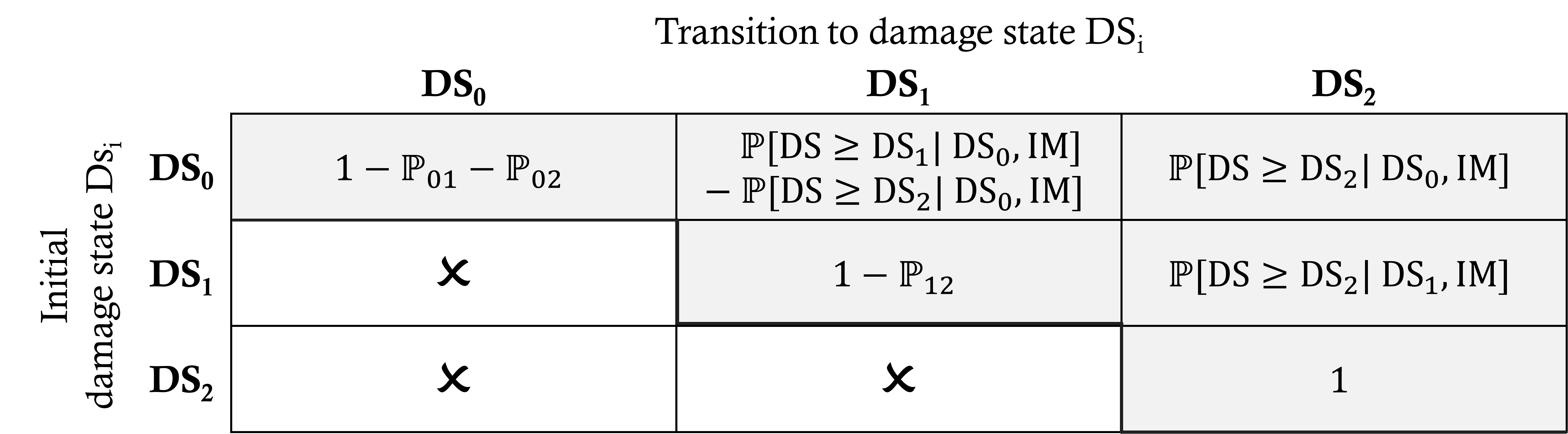}
		\caption*{(a)}
	\end{subfigure}
	\vfill
	\vspace{3mm}
	\centering
	\begin{subfigure}
		\centering
		\includegraphics[width=0.40\linewidth]{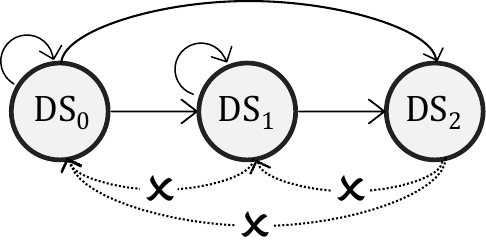}
		\caption*{(b)}
	\end{subfigure}
	\caption{Transition state (a) matrix and (b) diagram, respectively.}
	\label{fig:transitionStateGraph}
\end{figure}

\begin{figure*}[!ht]
	\centering
	\includegraphics[width=0.99\textwidth]{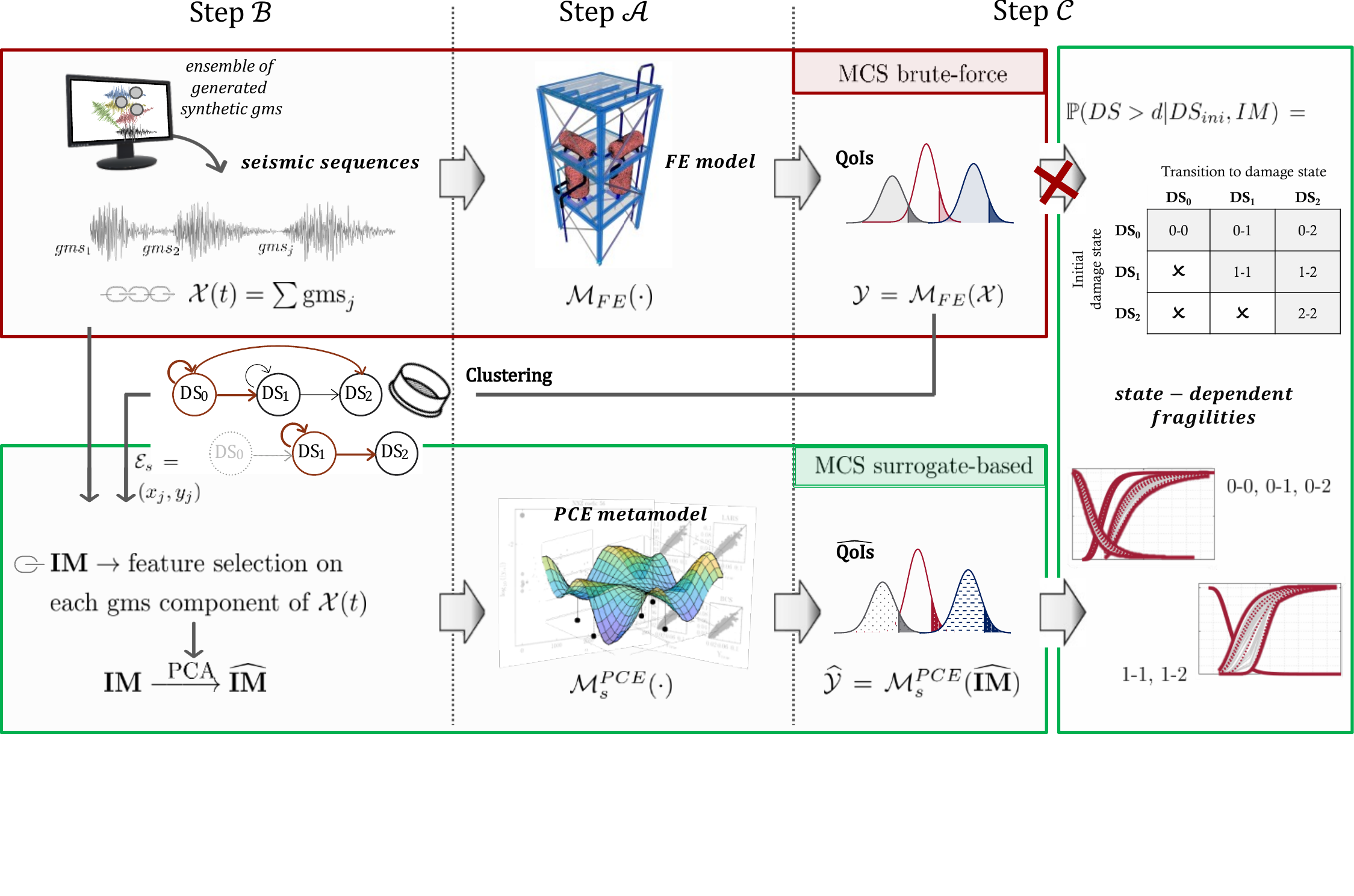}
	\vspace{-20mm}
	\caption{UQ-based framework and key steps for performing state-dependent fragility analysis. The top red-contoured row displays the brute-force MCS path on expensive-to-run NLTHAs (step $\mathcal{A}$ to $\mathcal{C}$), while
		the green one embeds the surrogate-based MCS method. Specifically, for the top row, complex and computationally demanding FE models are developed in step $\mathcal{A}$; whilst, in step $\mathcal{B}$, a GMM is deployed to compose stochastic seismic sequences to assign as input. However, the computational burden hinders the allowable number of MCS. This results in a constraint to the derivation of state-dependent fragility functions, as highlighted by the interrupted red-crossed arrow. Nevertheless, the obtained limited NLTHAs QoIs are clustered according to the damage initial state metric $\mathrm{DS}_{0,1,2}$. Those clustered data, coupled with the corresponding seismic event of the sequence, constitute the DoE $\mathcal{D}_s$ for the surrogate models (step $\mathcal{A}-\mathcal{B}$). First, a comprehensive vector of $\boldsymbol{IM}$s is extracted. Then, PCA is applied to limit the dimensionality of the input. This allows us to set PCE metamodels for each $\mathcal{D}_s$, $s \in \left[ \mathrm{DS}_0,\mathrm{DS}_1 \right]$. Finally, a vast number of surrogate-based MCS enables the derivation of state-dependent fragilities, (step $\mathcal{C}$).
	}
	\label{fig:heuristics}
\end{figure*}

State-dependent fragility curves are defined as a class of fragility curves conditioned not only by a measure of seismic intensity $\textbf{IM}$ (which in generalised form is a vector), but also by the initial state of (discrete) damage $\mathrm{DS_i}$ of the structure (see, \cite{bib:iervolino2015}). Hence, state-dependent fragilities enable the assessment of the vulnerability of a system that has already experienced damage, as defined by the generalised equation:
\begin{flalign}
	\mathbb{P} \displaystyle \Big[ \mathrm{DS_j} \vert \mathrm{DS_i}, \, \boldsymbol{IM} = \boldsymbol{im} \Big] \, &= \mathbb{P}  \Big[ \mathrm{DS} \geq \mathrm{DS_j} \vert \mathrm{DS_i}, \, \boldsymbol{IM} = \boldsymbol{im} \Big] \, - \mathbb{P} \, \Big[ \mathrm{DS} \geq \mathrm{DS_{j+1}} \vert \mathrm{DS_i}, \,\boldsymbol{IM} = \boldsymbol{im} \Big], \label{eq:state-dep} 
\end{flalign}
for $j > i$, and $i$ indices ranging among the identified and most severe damage limit states.
Figure~\ref{fig:transitionStateGraph}(a) shows the generalised transition probability state matrix for a system with three possible levels of damage, i.e., $\mathrm{DS_0}$ to $\mathrm{DS_2}$. 
Each row of the matrix represents the initial damage state or level; whilst each column indicates the final damage state. Each bin represents the transition probability between the initial (row) and final (column) state, including the permanence within the same level. 
The lower triangular part of the matrix represents the recovery processes of the investigated system, which, for brevity, we do not consider in this paper. 
Moreover, the ultimate damage limit state $\mathrm{DS_2}$ is considered an absorption state, i.e., a condition that, cannot be exited or recovered once reached. A typical example would be the collapse of the structural system. 
Similarly, Figure~\ref{fig:transitionStateGraph}(b) shows the Markovian diagram underlying the transition state matrix, e.g., the allowable jumps between the three-level damage states. 
\FloatBarrier
State-dependent fragility analysis requires a vast number of NLTHAs because each position of the transition matrix needs to be sufficiently populated.
As a consequence, time histories of multiple seismic events, e.g., seismic sequences, are cast and applied as input for the NLTHAs. This enables us to cover each transition state and different damage initial configurations effectively.
Then, the structural and non-structural system performances are clustered according to the damage reference metric. Hence, either empirical or parametric fragility functions are derived, conditioned on the initial damage state condition and the $\textbf{IM}$ of the seismic input. Figure~\ref{fig:heuristics} describes the fundamental steps of the workflow, i.e., from the FE and seismic input models definition (steps $\mathcal{A-B}$) to the complete NLTHAs and fragility computations (step $\mathcal{C}$). 
However, when considering realistic computational models, the significant computing demands of an extensive set of sequential NLTHAs generally hinder this direct derivation of state-dependent fragilities, as in step $\mathcal{C}$ upper-right corner of Figure~\ref{fig:heuristics}. To tackle this limit, the global UQ framework developed in \cite{bib:sudret2007} and \cite{bib:sudret2008} is adapted in the proposed methodology. The following three steps are used to define the UQ problem: (i) step $\mathcal{A}$, i.e., the definition of the computational model $\mathcal{M}(\cdot)$; (ii) step $\mathcal{B}$, description of the input parameters; (iii) step $\mathcal{C}$, propagation of the uncertainties and processing of the quantities of interests (QoIs).
Specifically, in the proposed framework, for the steps $\mathcal{A-B}$, we first perform a given number of sequential NLTHAs: 
\begin{flalign}
	\mathcal{Y} \, = \mathcal{M} \left( \ecs \right) = \mathcal{M}_{FE}\circ A(t, \mathcal{X}) =  \mathcal{M}_{FE}\left(A(t, \ecs)  \right), \label{eq:uq_fem} 
\end{flalign}
where $A(t,\ecs)$ is a given seismic sequence generated by a stochastic site-based ground motion model (GMM), and $\circ$ represents the function composition. $\ecs$ is a high-dimension random vector which represents both the epistemic (i.e., the stochastic nature of the model parameters), and the aleatory uncertainties (i.e., the Gaussian random variables representing the Gaussian noise). Specifically, $\ecs = \left[ \ecs_1, \dots , \ecs_j, \dots \ecs_M \right]^{T}$, where $\ecs_j$ is a random vector which represents the stochastic nature of a given ground motions (gms) and M the total number of seismic events that constitute the seismic sequence. Moreover, $\mathcal{M}_{FE}(\cdot)$ represents the expensive-to-run FE model; $\mathcal{Y}$ a random vector that collects the QoIs of the case study. In particular, $\mathcal{Y} = \left[ \mathcal{Y}_1, \dots , \mathcal{Y}_j, \dots \mathcal{Y}_M \right]^{T}$, where $\mathcal{Y}_j$ is the time series response of the associated $\ecs_j$ random seismic event.
\FloatBarrier
Next, instead of expensive-to-run FE models, cost-effective metamodels for each initial damage level---i.e., rows of the transition matrix of Figure~\ref{fig:transitionStateGraph}(a)---are tailored on the physics-informed problem and trained on a much smaller dataset. In particular, the QoIs resulting from the NLTHAs are clustered according to the predefined damage initial states, i.e., the number of rows of the transition state matrix of Figure~\ref{fig:transitionStateGraph}(a). For clarity, only the allowable transition state jumps are sketched. The clustering determines the pairs of $\left( \textbf{x}_j, y_j \right)$, where $j$ identifies each gm of the seismic sequences. This step serves to define different designs of experiment (DoE) depending on the initial state of the structure. Specifically, we identify the different DoEs with $\mathcal{D}_{s}$, where $s$ identifies the initial state of the structure, i.e., $s \in \{\mathrm{DS}_0,\mathrm{DS}_1\}$.
In the following step, a low-dimensional input representation is used instead of the time series sequences to build a time-invariant surrogate model. Specifically, we used a vector of $\textbf{IM}_j$, following the classical approach of vector-PSHA analysis. 
\noindent Besides the most popular IMs such as PGA and PGV, novel and less widespread IMs have also been introduced. Among them: the $\mathrm{I_{RG,a}}$ Riddell–Garcia Intensity acceleration and velocity measures, which minimise dispersion of hysteretic energy-dissipation spectra of inelastic systems; the $\mathrm{I_{F}}$ Fajfar Intensity, a compound IM that takes into account the damage capacity of medium-period structures; and the $\operatorname{\mathrm{E-ASA_{R,x}}}$ equipment relative average spectral acceleration, introduced in~\cite{bib:debiasio2015}. The last one is of particular interest since it allows us to consider shifts in the frequency range of the structure due to damage experienced by the installed equipment. The Table~\ref{tab:ch3_feature_IM} enlists the considered 41 IMs. Since among the list of IM several of them are strongly correlated, we seek specific patterns that allow dimensionality reduction. Therefore, we perform principal component analysis (PCA), to obtain a low dimension vector $\widehat{\textbf{IM}}_j$, which can be interpreted as a vector of uncorrelated pseudo-IMs.
\newcommand{\diff}{\mathrm{d}}
\newcommand{\dN}[1]{\diff{#1}} 

\begin{table*}[!ht]
	\centering
	\caption{A comprehensive list of ground motion IM parameters \cite{bib:debiasio2015} - \cite{bib:ardebili2016}.}{
		\label{tab:ch3_feature_IM}%
			\renewcommand{\arraystretch}{1.5} 
			\scalebox{0.8}{
    			\begin{tabular}{clccl}
    				\toprule
    				No.   & Description of IM & Symbol & Units & Mathematical model \\
    				\midrule
    				1     & Peak ground acceleration & $\mathrm{PGA}$ & $[\mathrm{m/s^2}]$ & $\max(\vert \ddot{u}(t) \vert)$ \\
    				2     & Peak ground velocity & $\mathrm{PGV}$ & $[\mathrm{m/s}]$ & $\max(\vert \dot{u}(t) \vert)$ \\
    				3     & Peak ground displacement & $\mathrm{PGD}$ & $[\mathrm{m}]$ & $\max(\vert u(t) \vert)$ \\
    				4,..,8 & Spectral displacement  & $\mathrm{Sd}$ & $[\mathrm{m}]$ & $S_d (T_{\textit{x}})$ \\
    				9,..,13 & Spectral velocity  & $\mathrm{Sv}$ & $[\mathrm{m/s}]$ & $S_v (T_{\textit{x}})$ \\
    				14,..,18 & Spectral acceleration  & $\mathrm{Sa}$ & $[\mathrm{m/s^2}]$ & $S_a (T_{\textit{x}})$ \\
    				& \multicolumn{3}{l}{\textit{$^*$ where $T^* \in \left[ 0.50; 0.35; 0.25; 0.15; 0.10 \right]$}} \\
    				19    & Arias intesity & $\mathrm{\mathrm{I_A}}$ & $[\mathrm{m/s}]$ & $\pi/2g \cdot \int_{0}^{t_f} a^2(t)\mathrm{d}T$ \\
    				20    & Total cumulative energy & $\mathrm{E_{cum}}$ & $[\mathrm{m^2/s^3}]$ & $\int \ddot{u}^2(t) \dN{t}$ \\
    				21    & Riddell–Garcia Intensity Acceleration & $\mathrm{I_{RG,a}}$ & $[\mathrm{m/s^{5/3}}]$ & $(PGA)^{+1}\cdot (T_d)^{+1/3}$ \\
    				22    & Riddell–Garcia Intensity Velocity & $\mathrm{I_{RG,v}}$ & $[\mathrm{m^{2/3}/s^{1/3}}]$ & $(PGV)^{+2/3}\cdot (T_d)^{+1/3}$ \\
    				23    & Significant time duration & $\mathrm{T_d}$ & $[\mathrm{s}]$ & $t_{95}-t_{05}$ \\
    				24    & Root mean square of acceleration & $\mathrm{RMS}$$\left(\ddot{u}\left(t\right)\right)$ & $[\mathrm{m/s^2}]$ & $ \sqrt{(\frac{1}{N}\sum_{n=1}^{N}\vert x_{N}^{2} \vert)}$ \\
    				25    & Characteristic intensity & $\mathrm{IC}$ & $[\mathrm{m^{3/2}/s^{3/2}}]$ & $\mathrm{RMS}(\ddot{u}(t))^{1.5}\cdot T_d^{0.5}$ \\
    				26    & Cumulative absolute velocity & $\mathrm{CAV}$ & $[\mathrm{m/s}]$ & $\int_{0}^{t_f}\vert a(t)\vert \dN{T}$ \\
    				27    & Cosenza–Manfredi Intensity & $\mathrm{I_{CM}}$ & $[\mathrm{-}$ & $2g/\pi \cdot (PGA)^{-1} (PGV)^{-1} (AI)^{+1}$ \\
    				28    & Average spectral acceleration & $\mathrm{ASA_{40}}$ & $[\mathrm{m/s}]$ & $2.5/f_1 \int_{0.6\cdot f_1}^{f_1} S_a(f,\varepsilon) \dN{f}$ \\
    				29    & Acceleration spectral intensity & $\mathrm{ASI}$ & $[\mathrm{m/s^2}]$ & $ \int_{0.1}^{0.5} S_a(T,\varepsilon)\dN{T}$ \\
    				30    & Effective peak acceleration & $\mathrm{EPA}$ & $[\mathrm{m/s^2}]$ & $1/2.5  \int_{0.1}^{2.5} S_a(T,\varepsilon)\dN{T}$ \\
    				31    & Velocity to acceleration ratio & $\mathrm{I_{v/a}}$ & $[\mathrm{s}]$ & $PGV/PGA$ \\
    				32    & Fajfar Intensity & $\mathrm{I_{F}}$ & $[\mathrm{m/s^{3/4}}]$ & $(PGV)^{+1}\cdot (T_d)^{+1/4}$ \\
    				33    & Mean frequency & $\mathrm{F_m}$ & $[\mathrm{1/s}]$ & $\sum_{i} U_i^2(f_i)/\sum_{i} U_i^2$ \\
    				34    & Rate of change mean frequency & $\mathrm{\dot{F}_m}$ & $[\mathrm{-}]$ & $\dN{F_m(T)} / \dN{t}$ \\
    				35    & Fourier amplitude spectrum area & $\mathrm{FAS_{area}}$ & $[\mathrm{m/s^2}]$ & $ \frac{1}{4 df}  \int_{f_1 - 2 df}^{f_1 + 2 df} U(f) \dN{f}$ \\
    				36,..,41 & Equipment relative average spectral acceleration & $\operatorname{\mathrm{E-ASA_{R_xx}^{**}}}$ & $[\mathrm{m/s^2}]$ & $  \frac{1}{f_1 \cdot (X_f - 1)} \cdot  \int_{f_1}^{X_f \cdot f_1} S_a(f,\varepsilon) \dN{f} $ \\
    				& \multicolumn{4}{l}{$^{**}$ \textit{$R$ indicates the chosen percentage of drop of the fundamental frequency ($X_f=1-(R/100)$);}} \\
    				& \multicolumn{3}{l}{$ R \in \left[ 40; 67; 80; 100; 150; 200 \right]$} \\
    				\bottomrule
    			\end{tabular}%
	}}
\end{table*}%
\noindent Particularly, PCA was used to select the least number of principal components (PCs) to satisfactorily characterise $\widehat{\textbf{IM}}$.
Finally, we build PCE surrogate models based on the pairs $(\widehat{\textbf{\textit{im}}}_j;y_j)$:
\begin{flalign}
	\widehat{\mathcal{Y}} \, =  \mathcal{M}^{PCE}_{s} \left( \widehat{\boldsymbol{IM}}~ \right), \label{eq:uq_PCE} 
\end{flalign}
where $\mathcal{M}^{PCE}_{s}(\cdot)$ is the PCE surrogate model; $\widehat{\mathcal{Y}}$ is the vector collecting the surrogated QoIs; $s \in [\mathrm{DS}_0;\mathrm{DS}_1]$ identifies the initial state of the structure. The vast number of MCS surrogate-based analyses enables us to derive non-parametric state-dependent fragilities functions, as defined in Eq.~\ref{eq:state-dep}.
\noindent Moreover, notice that the framework is not intrusive, meaning that the complex FE model is completely decoupled from the UQ analysis. This allows FE experts to work independently from UQ experts.
In the following Sections, the proposed UQ-based framework is applied twice. Specifically, in Section~\ref{ch3:benchmark} on an inexpensive-to-run benchmark case study, to validate the methodology and to illustrate its key steps and tools. Then, in Section~\ref{ch4:application}, we derive state-dependent fragility functions for a vertical tank installed on the industrial mock-up of the SPIF project. 
\newcommand{\apo}{\textquotesingle}
\section{Benchmark case: Hysteretic MDoF system} \label{ch3:benchmark}
To test the validity of the proposed framework, we examine the case of an equivalent mechanical cheap-to-evaluate multiple-degree-of-freedom (MDoF) shear-type system with Bouc-Wen hysteresis. This allows us: (i) to perform a vast number of sequential NLTHAs, and (ii) to evaluate state-dependent fragilities in terms of the inter-storey drift ratio (ID-ratio).
Then, the MCS brute-force fragilities are quantitatively and qualitatively compared with the surrogate-based MCS fragilities, derived by applying the UQ-based framework of Section~\ref{ch2:methodology}. 
\subsection{Step $\mathcal{A}$ - Computational model description}
\begin{figure}[!ht]
	\centering
	\begin{subfigure}{
			\hspace{8mm}
			\includegraphics[width = 0.20\linewidth]
			{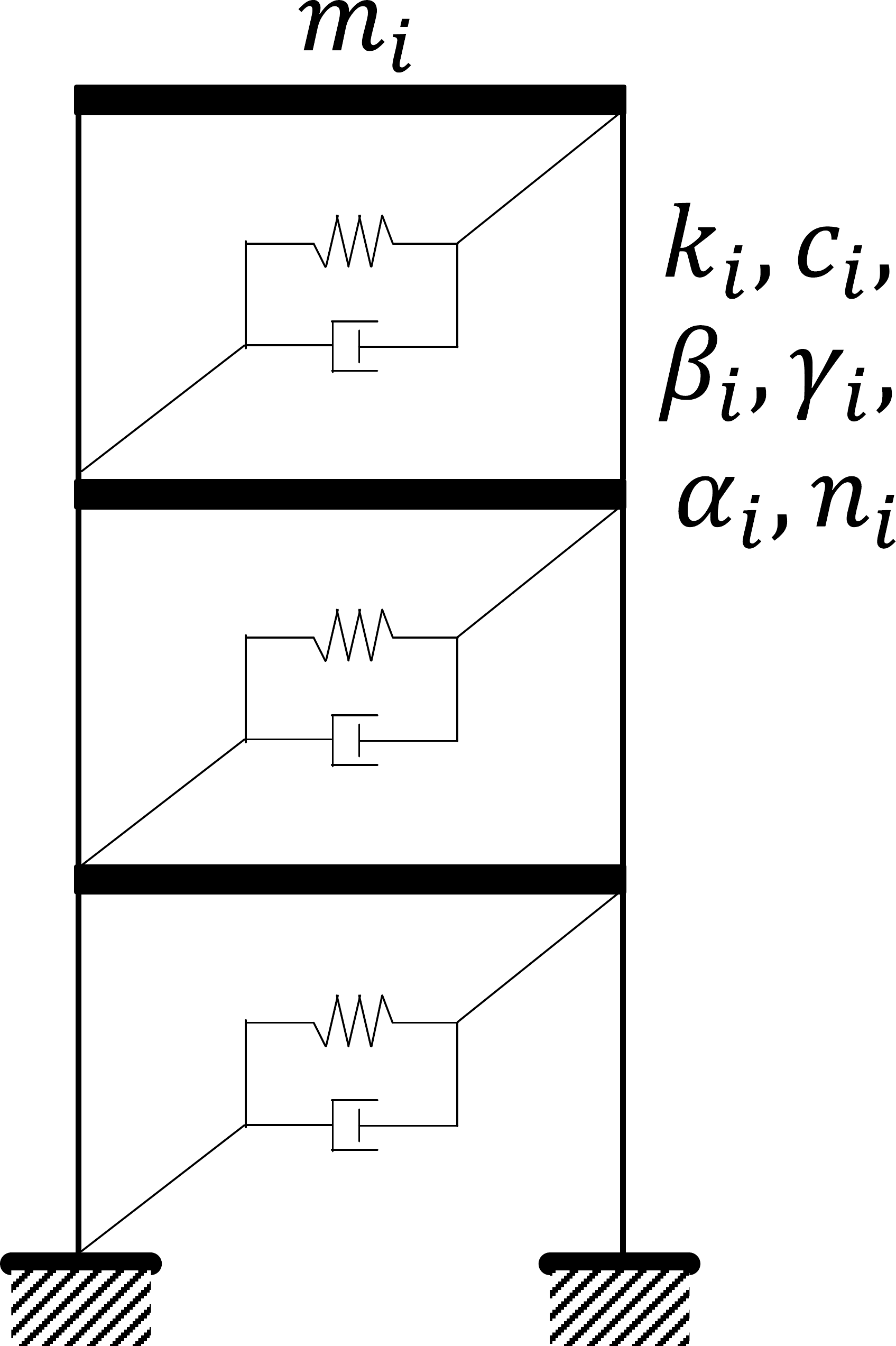}}
		\vspace{-6mm}
		\caption*{(a)}
	\end{subfigure}
	\begin{subfigure}
		\centering
		\includegraphics[ width = 0.50\linewidth]{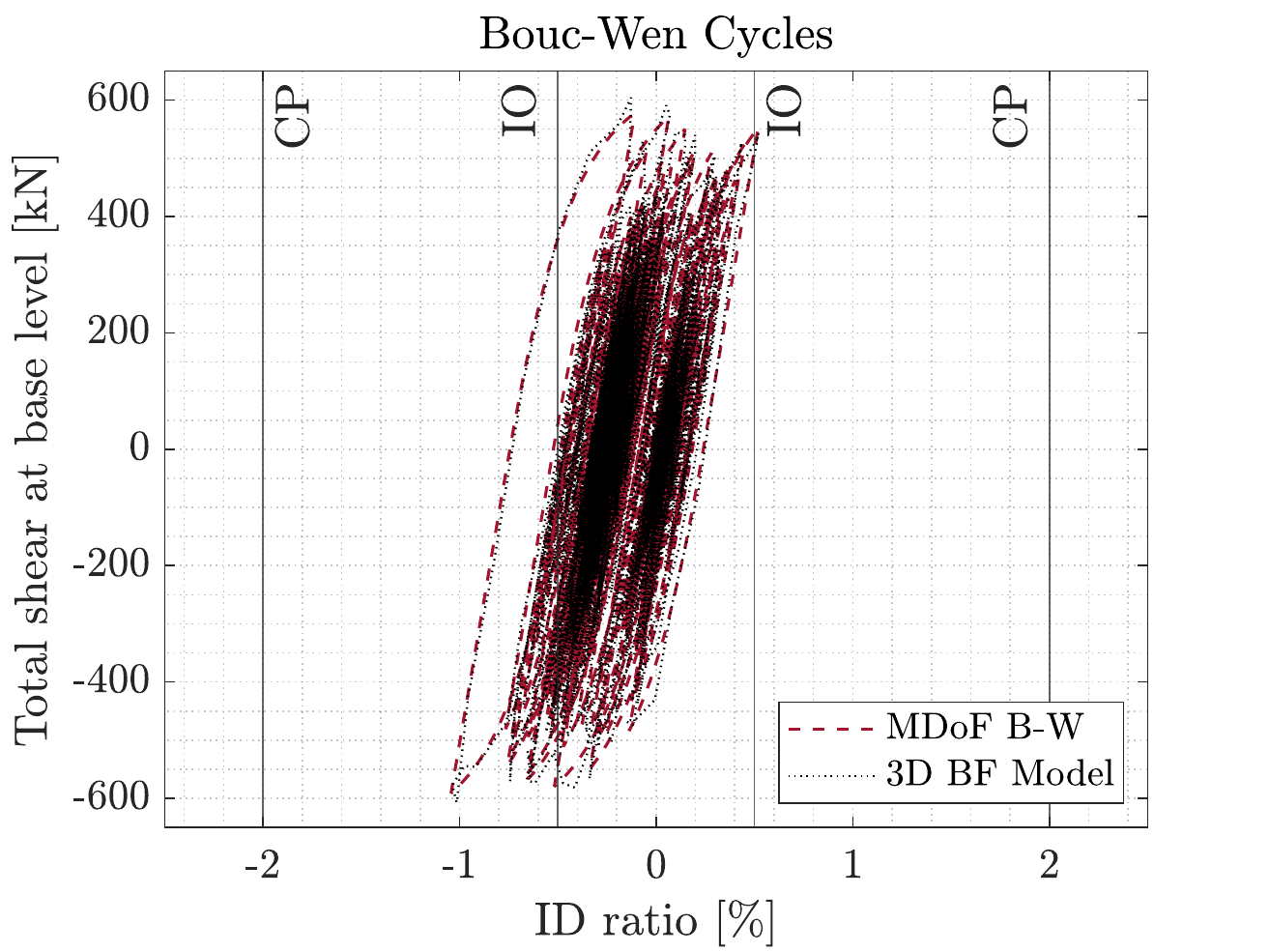}
		\vspace{-5mm}
		\caption*{(b)}
	\end{subfigure}
	\vspace{-3mm}
	\caption{(a) Schematics of the 2D MDoF system; (b) comparison of the hysteretic cycles at the top level of the 2D MDoF and the 3D BF model.}
	\label{fig:ch3_1_store_bw_cycle}
\end{figure}

According to the scheme depicted in Figure~\ref{fig:heuristics}, the MDoF system belongs to step $\mathcal{A}$ of the framework, i.e., the definition of the numerical model. 
The benchmark case study is a 2D condensation of the complex 3D industrial frame, namely SPIF \#2 system. Conversely, the full 3D system is studied in Section~\ref{sec:SPIF}. The SPIF \#2 project focused on investigating the seismic behaviour of a prototype multi-storey BF structure equipped with complex secondary industrial components by means of shaking table tests. More details on the project can be found in \cite{bib:quinci2023}, whilst the shake table test campaign and system performance are described in \cite{bib:nardin2022}.
The primary objective of the MDoF model is to efficiently capture the displacement and shear response time histories at each floor level of the complex BF system. 
To allow the execution of a significant number of NLTHAs, a strong emphasis on minimizing the computational resources and time is devoted. 
In this context, the engineering demand parameter (EDP) is represented by the maximum ID-ratio. 
Thus, according to Table~C1-3 \textit{Structural Performance Levels and Damage} of FEMA~356 \cite{bib:fema356}, the identified damage metrics $\mathrm{DS_i}$ are $0.5\%$ and $2.0\%$ of the ID-ratio, for Immediate Occupancy (IO) and Collapse Prevention (CP) limit states, respectively.
The numerical model implemented in-house with MATLAB\textsuperscript{\textcopyright} is defined by the semi-discretised system of equations of motion,
\begin{flalign}
	\textbf{M} \, \ddot{u}  \, + \,  \textbf{C} \, \dot{u}  \, + \, \textbf{K} \, \left[ \alpha u \, + \, \left( 1-\alpha \right) z \right]  = \,   -  \textbf{M} \, \ddot{u}_{g},
	\label{eq:motion}
\end{flalign}
where the stochastic base input $\ddot{{u}}_{g}$ is a realization of the process $A\left( t , \ecs  \right)$, later on presented in Eq.~\ref{eq:gmm} of Section~\ref{sec:input-gmm}. Both the mass $\textbf{M}$ and the stiffness $\textbf{K}$ matrices are calculated from the experimental data collected during the SPIF~\#2 test campaign~\cite{bib:nardin2022}; while, the damping $\textbf{C}$ is set to match a viscous damping ratio of 4.5$\%$, as detailed in ~\cite{bib:quinci2023}. The $\{ \alpha, \beta_N, \gamma_N, \delta D, \delta\nu, \delta\eta, z \}$ terms related to the Bouc-Wen hysteresis follow the formulation presented in \cite{bib:haukaas2003}, summarized as:
\begin{flalign}
	\dot{z} \, = \, \frac{\textbf{D} \, \dot{u} \, - \, \displaystyle \{ \beta  \vert \dot{u} \vert  z \vert z \vert^{n-1} \, + \, \gamma \vert z \vert^{n} \dot{u} \} \cdot \nu }{\eta}.
	\label{eq:bw}
\end{flalign}
All system parameters are reported in Table~\ref{tab:ch3_bw_par}, whilst Figure~\ref{fig:ch3_1_store_bw_cycle} depicts the comparison of the hysteresis at the top level for the reduced MDoF Bouc-Wen-based model versus the full FE BF model under seismic excitation.
\begin{table}[!ht]
	\centering
	\caption{Bouc-Wen parameters for the MDoF system.}{
		\label{tab:ch3_bw_par}%
		\scalebox{0.85}{
			\begin{tabular}{ccc}
				\toprule
				\multicolumn{1}{l}{\textbf{Parameter}} &       & \multicolumn{1}{c}{\textbf{Value}}  \\
				\midrule
				$\alpha$, $n$, $\beta_{N}$,$\gamma_{N}$ & [-]   & [0.01; 1.5;0.167; 0.50]   \\
				$\delta D$,$\delta \nu$,$\delta \eta$ & [-]   & [0.002; 1.00; 1.00]E-08 \\
				$K_0$ & [N/m] & [1.72;1.77;0.96]E+07 \\ $M$   & [kg]  & [9.5;14.5;12.4]E+03 \\
				\bottomrule
			\end{tabular}%
	}}
\end{table}%
\FloatBarrier
\subsection{Step $\mathcal{B}$ - Input definition}\label{sec:input-gmm}
Sequential chains of seismic events derived from the site-based GMM of \cite{bib:ADK2010} are used as the excitation. We consider the far-field GMM scenario described in \cite{bib:nardin2022} for the experimental test campaign of SPIF \#2. Briefly, the GMM is based on a modulated and filtered discretized white-noise process described by the following equation:
\begin{flalign}
	A\left( t , \ecs  \right) \, = \, q \left(t,\boldsymbol{\Theta}_q \right) \left[ \frac{\sqrt{2\pi S \Delta t}}{\sigma_{h}(t)} \displaystyle \sum_{l=1}^{k} h\left[ t-t_l, \boldsymbol{\Theta}_h(t_l)\right] \cdot Z_l \right] \mathrm{\ with}  \hspace{3mm} t_k \leq t < t_{k+1},
	\label{eq:gmm}
\end{flalign}
where $q(t,\boldsymbol{\Theta}_q)$ is the modulating function, $Z_l$ denote the standard normal Gaussian Random Variables, and  $ \sigma^2_h = 2 \pi S \Delta t \sum_{l=1}^{k} h^2[ t-t_l, \theta(t_l)]$ the standard deviation of the discrete filtered white-noise process. Moreover, $t_l$ is a set of equally spaced time points (with $l=0,1,..., L$, $t_0=0$, and $t_L$ representing the total duration of the motion) and $k = \text{int}(t_L/\Delta t)=0,1,\dots,L$.
As shown in \cite{bib:ADK2010}, $\boldsymbol{\Theta}_q \, = [I_a, t_{mid}, D_{5-95}]$ are the parameters of the modulating function, while $\boldsymbol{\Theta}_h \, = [\omega_{mid}, \dot{\omega}, \zeta]$ are the filter parameters. Provided with this representation, 
$\boldsymbol{\ecs}$ (in Eq. \eqref{eq:uq_fem}) is written as follows $\boldsymbol{\ecs} =  \displaystyle  \left[ I_a,\, D_{5-95},\, t_{mid},\, \omega_{mid}, \, \dot{\omega}_{mid}, \, \zeta_f, Z_1, \dots, Z_{L} \right]$. Observe that $\boldsymbol{\ecs}$ includes both an aleatory component represented by the Gaussian random variables and an epistemic component represented by the randomized GMM parameters. 
The marginal distributions of the GMM parameters are summarized in Table~\ref{tab:ch3-GMM-dataset}. In \cite{bib:nardin2022}, these distributions were inferred from selected INGV and ITACA datasets based on the following assumptions: (i) distance fault-site $R> 10$[km]; (ii) moment magnitude $M_w> 5.5$; (iii) main shock seismic events only; (iv) strong motion intensities expressed in terms of $PGA > 0.075$[g]. The complete list of the selected natural records and their characteristics is provided in \cite{bib:nardin2022}. Then, the calibrated GMM is used to generate an ensemble of 10$^4$ simulated ground motions (gms).
Based on the following assumptions: (i) mainshock events characterized by long return periods, (ii) no recovery/restoring processes, and (iii) negligible degradation effects with respect to seismic damage, gms are randomly extracted from the generated synthetic gms ensemble to compose sequences of seismic time histories, as sketched in Figure~\ref{fig:heuristics}-Step~$\mathcal{B}$. Observe that we do not consider after-shock sequences, but only sequences of main shock events.
Next, we construct the dataset in the following way. We simulate $10^4$ gms and then we randomly generate a sequence of $10$ seismic events by randomly permuting the $10^4$ gms. Thus, the resulting artificial dataset comprises $10^3$ sequences of $10$ gms. Finally, we reproduce 100 of such datasets. In particular, the generation of the GMM parameters is sampled using the 100 predefined seeds (for reproducibility). Notice that the white noise is not reproducible.
The adoption of earthquake sequences allows for considering seismic damage accumulation through time, thus mimicking the effect of non-pristine initial conditions on the structure. 
\begin{table*}[!ht]
	\centering
	\caption{Probabilities density distributions of the parameters of the site-based GMM.}{
		\label{tab:ch3-GMM-dataset}%
		\scalebox{0.90}{
			
			\begin{tabular}{llllccc}
				\toprule
				\textbf{Model } &       & \textbf{Units} & \multicolumn{1}{c}{\textbf{Distribution}} & \boldmath{}\textbf{$\mu$}\unboldmath{} & \boldmath{}\textbf{$\sigma$}\unboldmath{} & \textbf{Distribution} \\
				\textbf{Parameters} &       &       &       &       &       & \textbf{Bounds} \\
				\midrule
				$I_a$ & Arias Intensity & [m/s] & Log-normal & -0.46 & 0.51  & $(0; +\infty)$ \\
				$D_{5-95}$ & Time interval of 95\% of the $I_a$ & [s]   & Log-normal & 2.21  & 0.23  & $(0; +\infty)$ \\
				$t_{mid}$ & Time interval of 45\% of the $I_a$  & [s]   & Log-normal & 1.698 & 0.21  & $(0; +\infty)$ \\
				$\omega_{mid}/2\pi$ & Filter frequency at $t_{mid}$ & [Hz]  & Uniform & 4.8   & 1     & $[3.8; 5.8]$ \\
				$\zeta_f$ & Filter damping ratio & [-]   & Uniform & 0.35  & 0.1   & $[0.25;0.45]$ \\
				\bottomrule
				\multicolumn{7}{r}{\textit{*$\dot{\omega}_{mid}$, rate of change of frequency with $t$, is assumed constant and equal to -0.5}  }\\
			\end{tabular}%
	}}
\end{table*}%
\subsection{Step $\mathcal{C}$ - QoI response}
\subsubsection{Brute-force MCS for state-dependent fragility functions}
\begin{figure*}[!ht]
	\centering
	\includegraphics[width=0.99\linewidth,trim={0 0 0 1.5cm},clip]{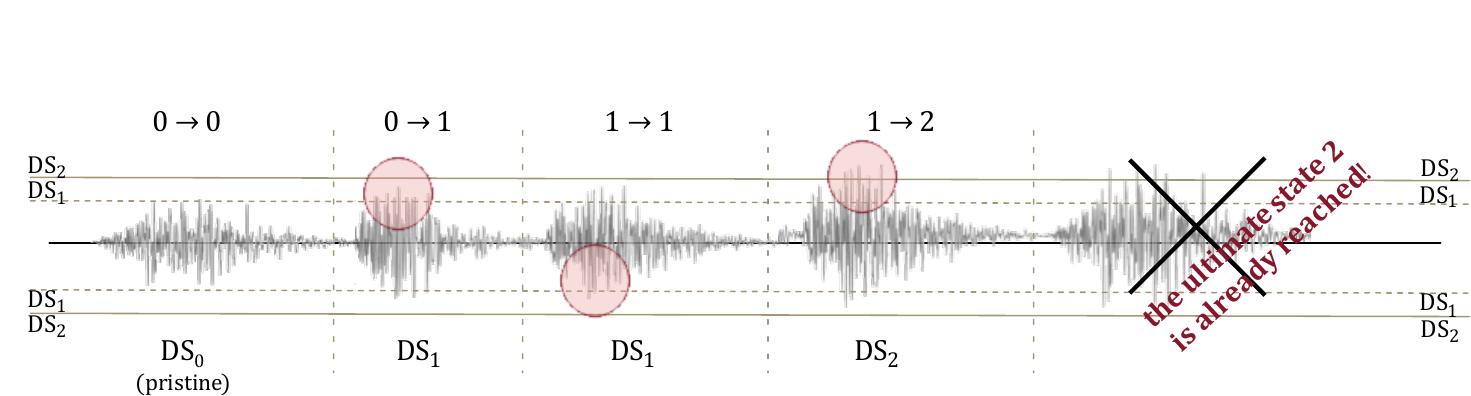}
	\caption{Example of seismic sequence time history response clustered according to the initial and final damage state level, e.g., ranging from $\mathrm{DS_0}$ (pristine) to $\mathrm{DS_2}$ (collapse).}
	\label{fig:ch3_transition_cluster}
\end{figure*}

\begin{figure*}[!ht]
	\centering
	\begin{tabular}{cc}
		\includegraphics[scale = 0.45]{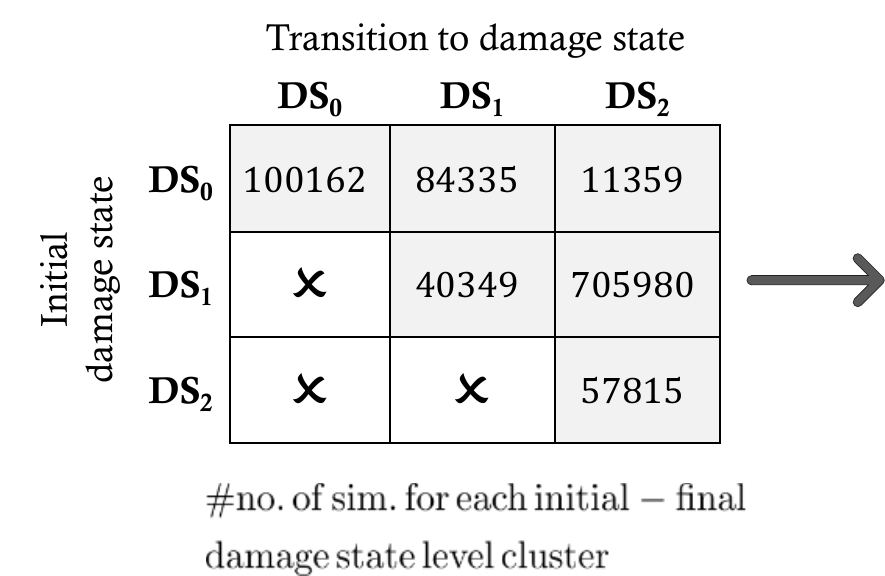} &
		\includegraphics[scale = 0.44]{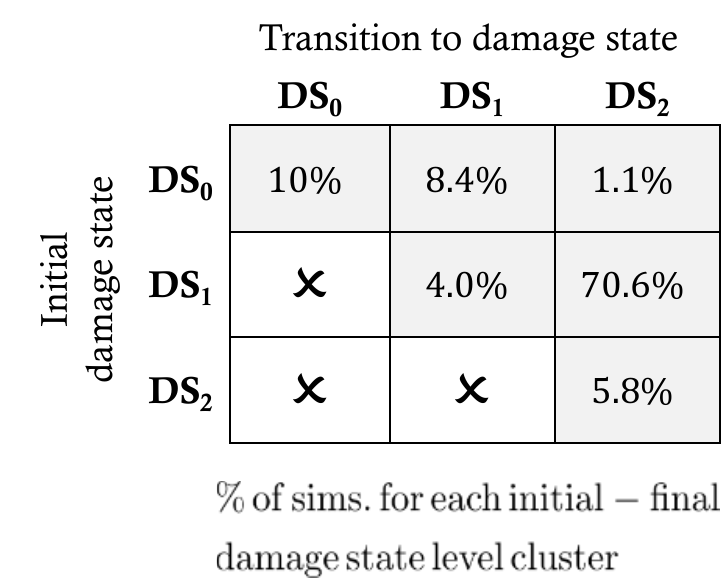} \\
		\hspace{5mm} (a) &
		\hspace{15mm} (b) \\
	\end{tabular}
	\caption{Transition state matrices: (a) counters and (b) percentages of simulation for each initial-final damage state level cluster, respectively.}
	\label{tab:ch3_cluster_mdof_trx}
\end{figure*}

\begin{figure*}[!ht]
	\centering
	\includegraphics[scale=0.3, trim = {0.5cm 0cm 0.5cm 0cm},clip]{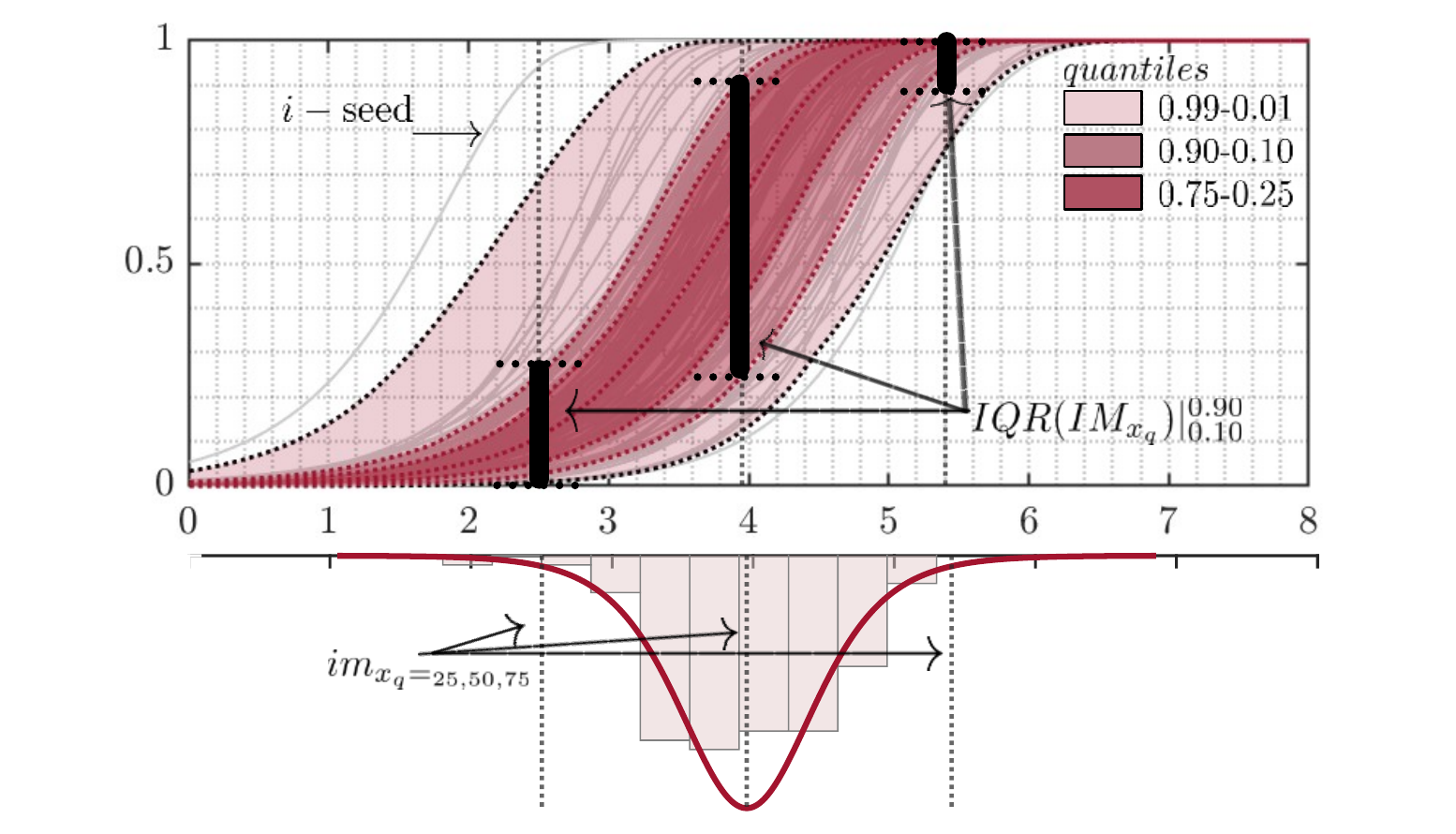}
	\caption{$\mathrm{\beta_{eff}}$ criterion explained: step (i), determine the  $x_q \, = \, 25^{th}-50^{th}-75^{th}$ quantiles on the marginal IM; step (ii), evaluate the $\mathrm{IQR}|_{0.10}^{0.90}$ on $x_q$; step (iii), repeat $\forall IM_i$ and find the minimum.}
	\label{fig:ch3_beta_index_explained}
\end{figure*}

\begin{figure*}[!ht]
	\centering
	\includegraphics[width=0.99\columnwidth]{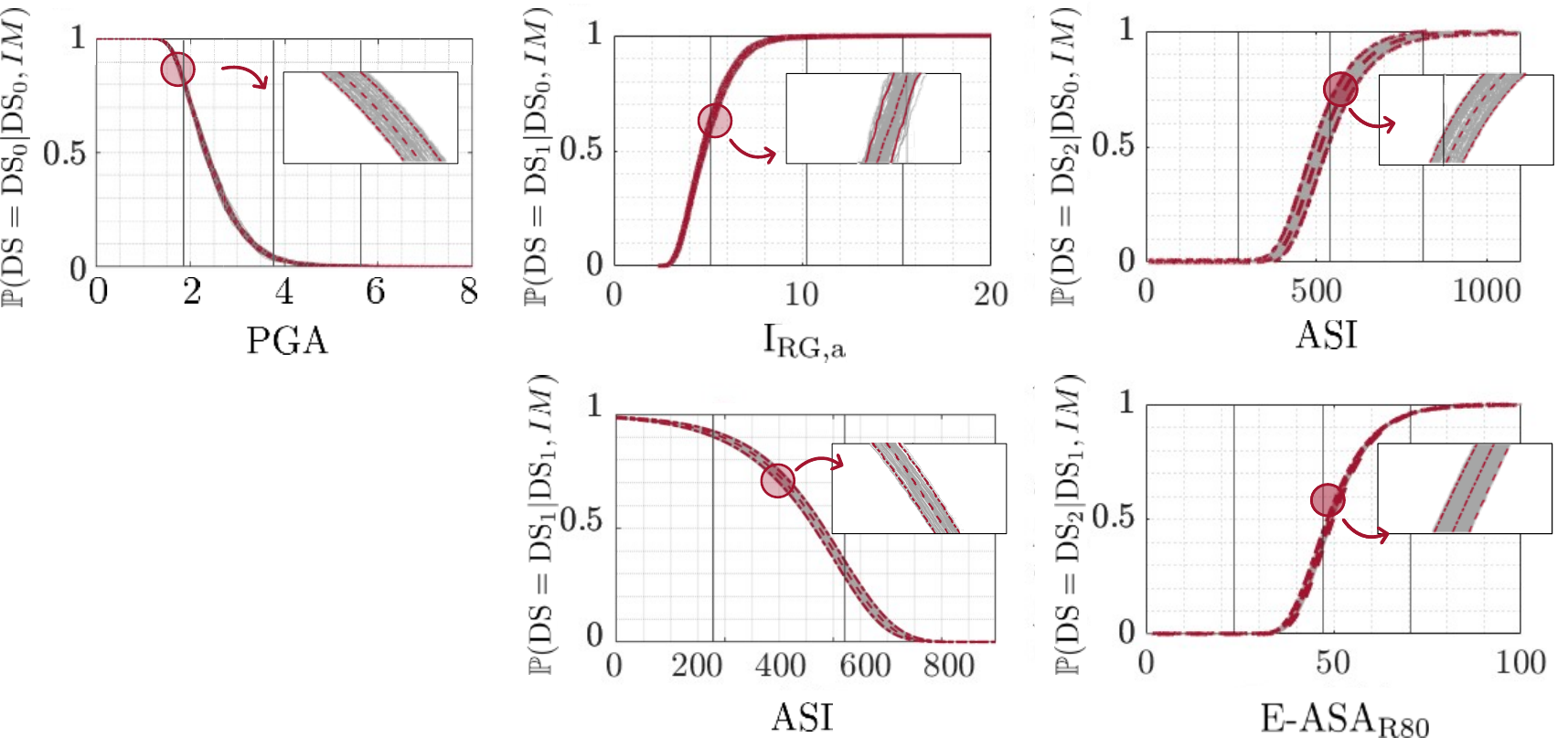}  
	\caption{Brute-force MCS state-dependent fragility functions: grey lines represent single seed simulations; dashed-dot dark-red lines depict the $90\%$ confidence bounds, whilst dashed-dark red ones the $50\%$. Different IMs are adopted in $x$-axes, according to the optimal $\mathrm{\beta_{eff}}$ for each transition state.}
	\label{fig:ch3_mc_fragility}
\end{figure*}

\begin{figure*}[!ht]
	\centering
	\includegraphics[width=0.99\linewidth]{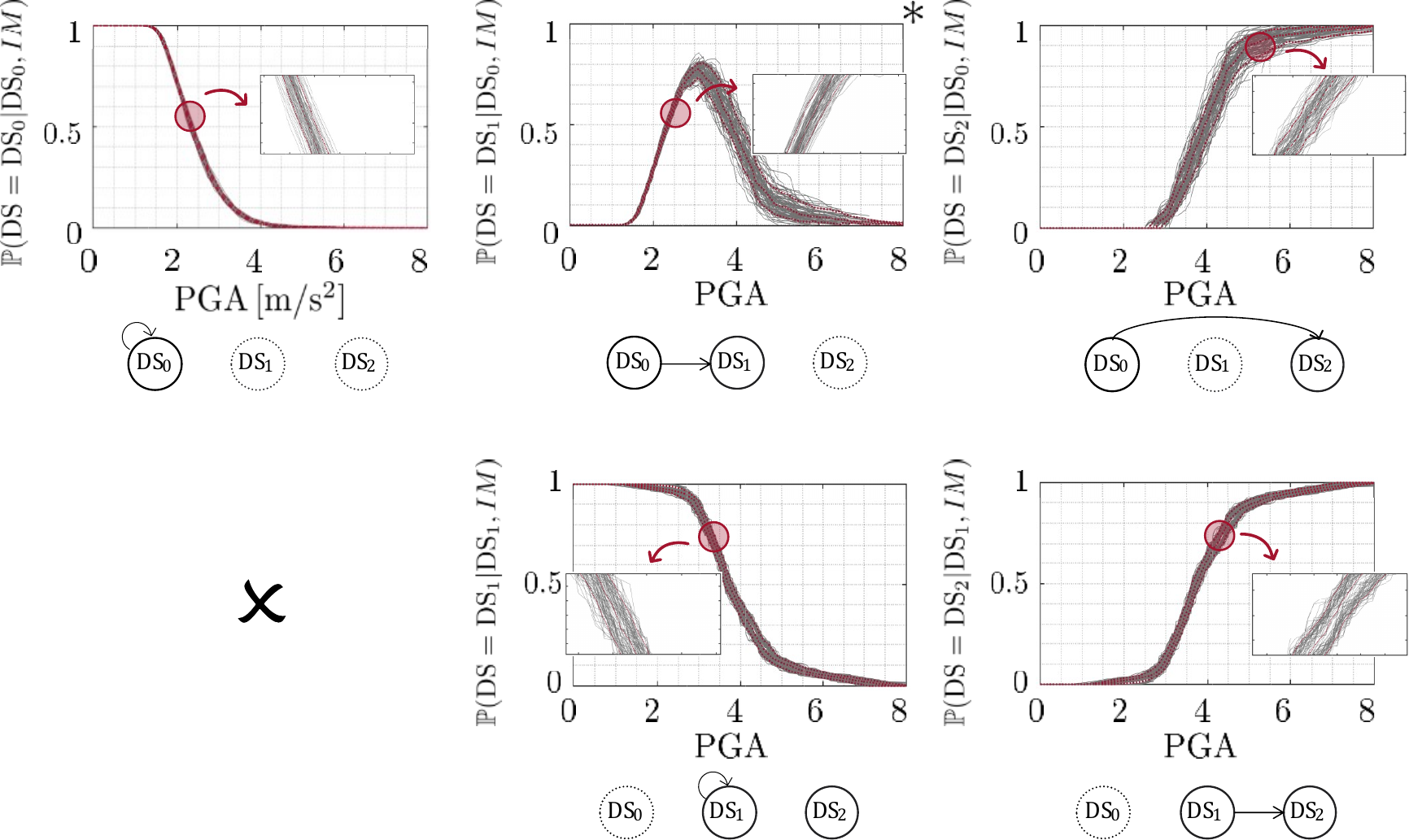}
	\caption{Brute-force MCS state-dependent fragility functions, by means of $\mathrm{PGA}$: grey lines represent single seed simulations; dashed-dot dark-red lines depict the $99\%$ confidence bounds, whilst dashed-dark red ones the $50\%$. $\mathrm{PGA}$ is the IM adopted, according to the global optimal $\mathrm{\beta_{eff,glob}}$ among transition states. Notice that this Figure is in accordance with Eq.~\ref{eq:state-dep}; therefore, the subplot at line 1 column 2 represents $\mathbb{P}(\mathrm{DS}=\mathrm{DS_1}|\mathrm{DS_0},IM)$. Therefore, this is not a commonly used ``fragility function,'' which reports the exceedance probability, i.e. $\mathbb{P}(\mathrm{DS}\ge\mathrm{DS_1}|\mathrm{DS_0},IM)$. Figure~\ref{fig:ch3_state_01_PGA}, in \ref{app:B_MDOF_collection}, reports the commonly used fragility function.}
	\label{fig:ch3_mcs_fragility_pga}
\end{figure*}

\noindent A total amount of $10^6$ sequential NLTHAs were performed, thanks to the minimal computational effort required by a single sequential NLTHA\footnote{$\approx$3s on an Intel(R) Core(TM) i9-10900K CPU @3.70GHz, 10 Core(s) - 128 GB RAM}.
For each simulation, the initial and final damage configurations were identified based on the EDP. 
Next, simulations were clustered according to the initial and the final damage state level associated with each ground motion of the seismic sequence, as in Figure~\ref{fig:ch3_transition_cluster}. 
Thus, six combinations among initial-final damage configurations are possible: \textquotesingle0-0\textquotesingle ~, \textquotesingle0-1\textquotesingle ~, \textquotesingle0-2\textquotesingle ~, \textquotesingle1-1\textquotesingle ~, \textquotesingle1-2\textquotesingle ~, \textquotesingle2-2\textquotesingle. Those represent the transition states of the system, as sketched in Figure~\ref{fig:transitionStateGraph}(a). Specifically, the \textquotesingle0-0\textquotesingle ~, \textquotesingle0-1\textquotesingle ~, \textquotesingle0-2\textquotesingle ~ identify, given a pristine $\mathrm{DS_0}$ initial state condition, the permanence in the $\mathrm{DS_0}$ and transition to damage state level $\mathrm{DS_1}$ and $\mathrm{DS_2}$, respectively. Similarly, \textquotesingle1-1\textquotesingle ~,\textquotesingle1-2\textquotesingle ~ identify the permanence in $\mathrm{DS_1}$ and transition to damage state level $\mathrm{DS_2}$, given an initial damage state level $\mathrm{DS_1}$. Finally, the \textquotesingle2-2\textquotesingle ~ refers to the absorption state of collapse.
\FloatBarrier
Table~\ref{tab:ch3_cluster_mdof_trx}(a) collects the numbers of clustered results of the MCS, while Table~\ref{tab:ch3_cluster_mdof_trx}(b) the percentages. Nearly 74\% of the total simulations exhibited an initial damage level of $\mathrm{DS_1}$, with approximately 70\% concentrated within the \textquotesingle1-2\textquotesingle ~ cluster. Conversely, around 20\% of simulations were performed with no initial damage, e.g., $\mathrm{DS_0}$. The remaining ~6\% of simulations resulted in an absorption collapse state.
According to Eq.~\ref{eq:state-dep}, to evaluate state-dependent fragilities, an optimal IM, or a vector of optimal IMs, needs to be defined. To do so, we implemented an efficiency criterion $\mathrm{\beta_{eff}}$ based on \cite{bib:ardebili2016} and \cite{bib:zentner2017}. Based on Figure~\ref{fig:ch3_beta_index_explained}, first, we determine the $x_q \, = \, 25^{th}-50^{th}-75^{th}$ quantiles of each marginal  IM distribution of Table~\ref{tab:ch3_feature_IM}. Second, we evaluate the $90^{th}$ inter-quantile range $\mathrm{IQR}$ in correspondence with the previous quantiles of the marginal  IM distributions. Third, we sum these $90^{th} \, \mathrm{IQR}$ ranges. Finally, the optimal $\mathrm{\beta_{eff}}$ is the minimum among the $IM_i$. Eq.~\ref{eq:betaIndex} describes the criterion:
\begin{flalign}
	\mathrm{\beta_{eff}}(IM_i)  = \, \mathrm{min} \left( \sum_{x_q} \mathrm{IQR}_{x_q}(IM_{i})\vert_{0.10}^{0.90} \right), 
	\label{eq:betaIndex}
\end{flalign}
with $IM_{i}$ ranging across the 41 elements of the vector $\boldsymbol{IM}$ defined in Table~\ref{tab:ch3_feature_IM}. Next the optimal $IM$ is selected as follow $IM^* = \operatorname*{argmin}_{IM_i} \, \mathrm{\beta_{eff}}(IM_i)$.  This procedure is applied for each transition state. It follows that transitions can have different optimal $IM^*s$. The collection of the optimal transition $IM^*s$ defines the optimal vector $\boldsymbol{IM}^*$, which is used to construct state-dependent fragility functions.  
At this stage, brute-force MCS on cheap-to-evaluate systems allows us to reach a full probabilistic description of fragilities. Given the vast number of simulations, empirical fragilities were evaluated for each transition state. Specifically, non-parametric curve-fitting on the cumulative distributions of data against the optimal IM was implemented, as in~\cite{bib:gem2015}.
\FloatBarrier
Figure~\ref{fig:ch3_mc_fragility} represents the collection of state-dependent fragility functions based on $IM^*$. Notice that this Figure does not represent the transition matrix defined in Eq.~\ref{eq:state-dep}. In fact, this is used to highlight the components of $\boldsymbol{IM}^*= [PGA,\ I_{RG,a},\ ASI,\ E-ASA_{R80}]$.
Each one of the 100 MCS seeds is associated with a grey line. 
The curves representing the 90\% confidence bounds are displayed in a dashed-dotted dark red style, while those for the 50\% confidence bounds are represented with dashed lines. 
The first row, corresponding to a pristine initial damage state condition, is characterized mainly by PGA-derived $IM$s, while the second row is characterized by $\mathrm{S_a}$-related features. Furthermore, it becomes evident that the optimal $IM$ changes as moving toward more severe damage states or starting from a non-pristine initial condition. Particularly, the evolving $IM$ is characterized by features capable of capturing reductions in stiffness due to the accumulation of damage, and shifts in the frequency range of the structure, such as $\operatorname{\mathrm{E-ASA_{R*}}}$. The $\mathrm{\beta_{eff}}$ values are collected in Table~\ref{tab:ch3_mdof_mc_ranking}. 
\begin{table*}[htbp]
	\centering
	\caption{Brute-force MCS state-dependent fragility functions: top five $\mathrm{\beta_{eff}}$ efficiency indices for each transition state.}
	\label{tab:ch3_mdof_mc_ranking}%
	\scalebox{0.65}{
		\begin{tabular}{llrllrllrllrll}
			\cmidrule{1-2}\cmidrule{4-5}\cmidrule{7-8}\cmidrule{10-11}\cmidrule{13-14}    \multicolumn{1}{c}{\textbf{Frag. \apo0-0\apo}} & \multicolumn{1}{c}{\boldmath{}\textbf{$\mathrm{\beta_{eff}}$}\unboldmath{}} &       & \multicolumn{1}{c}{\textbf{Frag. \apo0-1\apo}} & \multicolumn{1}{c}{\boldmath{}\textbf{$\mathrm{\beta_{eff}}$}\unboldmath{}} &       & \multicolumn{1}{c}{\textbf{Frag. \apo0-2\apo}} & \multicolumn{1}{c}{\boldmath{}\textbf{$\mathrm{\beta_{eff}}$}\unboldmath{}} &       & \multicolumn{1}{c}{\textbf{Frag. \apo1-1\apo}} & \multicolumn{1}{c}{\boldmath{}\textbf{$\mathrm{\beta_{eff}}$}\unboldmath{}} &       & \multicolumn{1}{c}{\textbf{Frag. \apo1-2\apo}} & \multicolumn{1}{c}{\boldmath{}\textbf{$\mathrm{\beta_{eff}}$}\unboldmath{}} \\
			\multicolumn{1}{c}{\textbf{Optimal IM}} & \multicolumn{1}{c}{\textbf{index}} &       & \multicolumn{1}{c}{\textbf{Optimal IM}} & \multicolumn{1}{c}{\textbf{index}} &       & \multicolumn{1}{c}{\textbf{Optimal IM}} & \multicolumn{1}{c}{\textbf{index}} &       & \multicolumn{1}{c}{\textbf{Optimal IM}} & \multicolumn{1}{c}{\textbf{index}} &       & \multicolumn{1}{c}{\textbf{Optimal IM}} & \multicolumn{1}{c}{\textbf{index}} \\
			\cmidrule{1-2}\cmidrule{4-5}\cmidrule{7-8}\cmidrule{10-11}\cmidrule{13-14}    $\mathrm{PGA}$ & 9.60E-02 &       & $\mathrm{I_{RG,a}}$ & 1.83E-02 &       & $\mathrm{ASI}$ & 8.59E-02 &       & $\mathrm{ASI}$ & 7.00E-03 &       & $\operatorname{\mathrm{E-ASA_{R80}}}$ & 1.96E-02 \\
			$\mathrm{IC}$ & 1.01E-01 &       & $\mathrm{PGA}$ & 2.10E-02 &       & $\mathrm{EPA}$ & 9.55E-02 &       & $\mathrm{EPA}$ & 7.70E-03 &       & $\operatorname{\mathrm{E-ASA_{R200}}}$ & 2.24E-02 \\
			$\mathrm{I_{RG,a}}$ & 1.03E-01 &       & $\mathrm{I_{RG,v}}$ & 2.50E-02 &       & $\mathrm{RMS{(\ddot{u})}}$ & 9.82E-02 &       & $\mathrm{PGA}$ & 7.70E-03 &       & $\mathrm{E-ASA_{R100}}$ & 2.31E-02 \\
			$\operatorname{\mathrm{E-ASA_{R80}}}$ & 1.07E-01 &       & $\operatorname{\mathrm{E-ASA_{R80}}}$ & 2.62E-02 &       & $\mathrm{PGA}$ & 9.83E-02 &       & $\mathrm{I_{RG,v}}$ & 7.90E-03 &       & $\operatorname{\mathrm{E-ASA_{R150}}}$ & 2.63E-02 \\
			$\mathrm{RMS{(\ddot{u}})}$ & 1.10E-01 &       & $\mathrm{PGD}$ & 2.63E-02 &       & $\operatorname{\mathrm{E-ASA_{R80}}}$ & 1.17E-01 &       & $\mathrm{I_{RG,a}}$ & 8.10E-03 &       & $\mathrm{I_{RG,a}}$ & 4.10E-02 \\
			\cmidrule{1-2}\cmidrule{4-5}\cmidrule{7-8}\cmidrule{10-11}\cmidrule{13-14}    \end{tabular}%
	}
\end{table*}%
A limitation of this approach is that one needs to develop a fragility model based on $\boldsymbol{IM}^*$, which can become computationally demanding (in terms of memory allocation). In this specific example, the state-dependent fragility models, $\mathbb{P}(\mathrm{DS}_j|\mathrm{DS}_i,\boldsymbol{IM}^*)$ are four-dimensional functions. Moreover, a vector-based seismic risk analysis needs to be developed to use this model directly. While this is not a limit in a Monte-Carlo-based seismic risk analysis, it becomes prohibitive for direct integration. A more straightforward approach is to use state-dependent fragilities as functions of one optimal $IM$ for the entire transition states. This is especially convenient when performing risk assessment since it allows the use of directly available GMPEs.  Therefore, we proposed a global efficiency metric defined as 
\begin{flalign}
	\mathrm{\beta_{eff,glob}}(IM_i) \, = \, \mathrm{min} \left( \sum_{s=0}^{S} \frac{\mathrm{\beta_{eff}}(IM_i)\vert_s}{\left( \sum_j^{41} \mathrm{\beta_{eff}}(IM_j) \right)\vert_s}  \right), \hspace{5 mm} \forall IM_{i} \in \textbf{IM} \, \mathrm{and} \,  s \, = \{ 0 , \dots , S \}  \label{eq:global_beta}  
\end{flalign}
where $s$ represents the allowable transition states, e.g., \apo0-0\apo,\apo0-1\apo,\apo0-2\apo,\apo1-1\apo,\apo1-2\apo. Specifically, we first normalize the $\mathrm{\beta_{eff}}$ index for each transition state over the $i$ to 41 IMs. Then, we sum the same $IM_i$ across the transition states, and, finally, we select the minimum. Table~\ref{tab:global_beta} gathers the global ranking for optimal IMs.
\begin{table*}[htbp]
	\centering
	\scalebox{0.75}{
		\begin{tabular}{llc}
			\toprule
			& \multicolumn{1}{c}{\textbf{Global}} & \boldmath{}\textbf{$\mathrm{\beta_{eff,glob}}$}\unboldmath{} \\
			& \multicolumn{1}{c}{\textbf{Optimal IM}} & \textbf{index[\%]} \\
			\midrule
			1.    & $\mathrm{PGA}$ & 9.12 \\
			2.    & $\mathrm{PGD}$ & 9.20 \\
			3.    & $\operatorname{\mathrm{E-ASA_{R800}}}$ & 9.21 \\
			4.    & $\mathrm{I_{RG,a}}$ & 9.39 \\
			5.    & $\mathrm{Sa(T_1)}$ & 9.49 \\
			\bottomrule
	\end{tabular}}
	\captionof{table}{Ranking of the optimal IMs according to the global $\mathrm{\beta_{eff,glob}}$, defined in Eqn.~\ref{eq:global_beta}.} \label{tab:global_beta}
\end{table*}
\FloatBarrier
\vspace{-3mm}
\noindent Figure~\ref{fig:ch3_mcs_fragility_pga} shows the probabilistic description of state-dependent fragilities as functions of the global optimal IM, i.e., the $\mathrm{PGA}$.
According to the definition of Eq.~\ref{eq:state-dep}, Figure~\ref{fig:ch3_mcs_fragility_pga} reproduces the transition matrix of Figure~\ref{fig:transitionStateGraph}(a), i.e., the probabilities of transition to a specific $\mathrm{DS_i}$ state. Specifically, in the first row, the probabilities of permanence in $\mathrm{DS_0}$ given $\mathrm{DS_0}$, i.e., $\mathbb{P}_{00}$, is defined as $\mathbb{P}\left( \mathrm{DS} = \mathrm{DS_0} \vert \mathrm{DS_0}, IM\right) \, = \boldsymbol{1} -\mathbb{P}_{01} - \mathbb{P}_{02}$. Here, particular attention is paid to the definition of $\mathbb{P}_{01}$, as also highlighted in the Figure by the $^*$ character. Indeed, the $\mathbb{P}_{01}$ is defined as the transition probabilities to attain $\mathrm{DS_1}$ only. This is different than the probabilities of exceedance $\mathrm{DS_1}$, which implies the probabilities of attaining $\mathrm{DS_1}$ or worse, e.g., $\mathrm{DS_2}$, as reported for completeness and clarity in Figure~\ref{fig:ch3_state_01_PGA} of the \ref{app:B_MDOF_collection}. The transition diagrams below each fragility clarify this by remarking only the allowable transitions. Similarly, $\mathbb{P}_{02}$ is defined as the probability of reaching $\mathrm{DS_2}$. There are no other worse conditions, since $\mathrm{DS_2}$ stands for the collapse case. Note a straightforward definition for the second row, since only two states are allowed. Particularly, in this row, an $\mathcal{\boldsymbol{x}}$ marks the transition state associated with recovery processes, not addressed in this paper. Finally, the last row of Figure~\ref{fig:transitionStateGraph}(a) is not included because represents the absorption case, i.e., the unit probability of collapse given collapse as initial conditions.  
Moreover, Figure~\ref{fig:ch3_mcs_fragility_pga} reveals a greater dispersion in the \apo0-2\apo~ curves with respect to the others. This is strictly related to the lower number of clustered data in this transition state. Indeed, the dispersion of fragilities is positively related to the number of clustered data that describes the fragilities functions. 
\subsubsection{Surrogate-based MCS for state-dependent fragility functions}
\begin{figure}[!ht]
	\centering
	\includegraphics[width=0.40\textwidth]{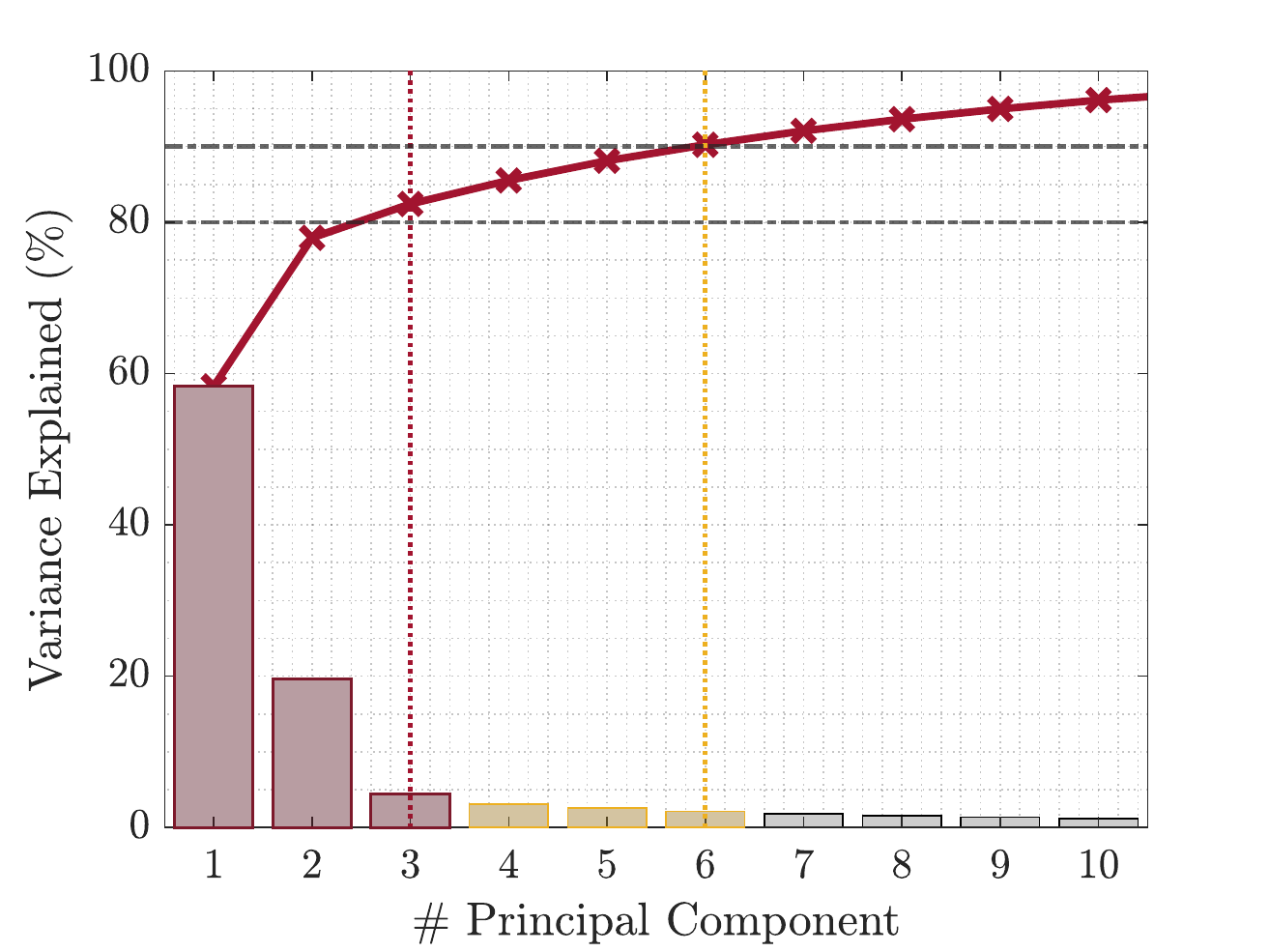}
	\caption{Scree plot of the seismic input.} 
	\label{fig:ch2_scree_plot}
\end{figure}

\begin{figure*}
	\begin{minipage}{0.88\textwidth}
		\begin{align}
			\vcenter{\hbox{ 
					\hspace{-10mm}\includegraphics[width=0.50\linewidth]{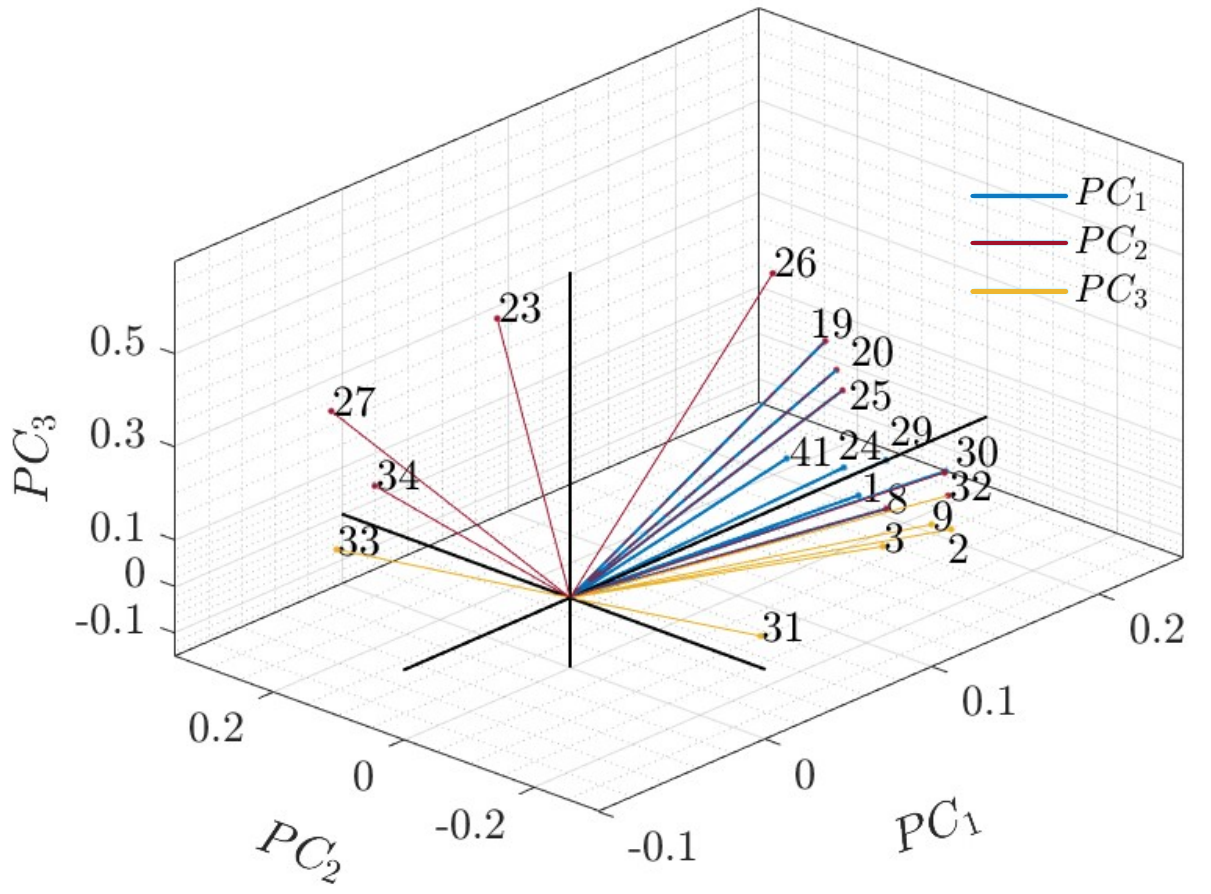}}
			} \hspace{3 mm}&
			\begin{aligned}
				PC_1 = \, &+0.202 \, \mathrm{ASI} \,
				+0.198 \,\mathrm{EPA} +0.190 \, \mathrm{\operatorname{E-ASA_{R_{200}}}}\, +  \\
				&+0.190\, \mathrm{I_{CM}} \, +0.186\, \mathrm{E_{cum}} \, +0.183\, \mathrm{PGA} \, + \\
				&+0.180\, \mathrm{I_A} \, +0.175 \, \mathrm{S_v(T_5)} \, +0.175 \, \mathrm{S_d(T_5)} \,  + \dots
				\\
				PC_2 = \, &-0.021 \mathrm{F_m} \, +0.005 \, \mathrm{AV} \, -0.047 \, \mathrm{I_{CM}} \, +0.136\, \mathrm{S_v(T_1)} +\\
				&+0.136 \, \mathrm{S_a(T_1)} \, +
				0.136 \, \mathrm{S_d(T_1))}\, 
				+0.109 \, \mathrm{PGD}  \, + \\
				& +0.146 \, \mathrm{I_{RG,v}} \, +0.150 \, \mathrm{PGV} \, +0.149 \, \mathrm{I_{F}} \, + \dots \\
				PC_3 = \, &-0.036 \, \mathrm{T_{d}} \, +0.149 \, \mathrm{CAV} \, -0.047 \, \mathrm{I_{CM}} \, + \, -0.061 \mathrm{F_m} \, +\\
				&+0.180 \, \mathrm{I_A} \, +0.187 \, \mathrm{E_{cum}} \, +0.146 \, \mathrm{I_{RG,v}} \, +0.190 \, \mathrm{I_c} \, +\\
				&+0.149 \, \mathrm{I_{F}} \, +0.181 \, \mathrm{S_d(T_5)} \,  + \dots 
				\notag
			\end{aligned}
		\end{align}
		\captionof{figure}{3D biplots of PCs 1, 2 and 3 for the IMs. For clarity, only the relevant $im_i$, identified by numbers as labelled in Table~\ref{tab:ch3_feature_IM}, are depicted.}
		\label{fig:ch2_biplot}
	\end{minipage}
\end{figure*}

\begin{figure*}[!ht]
	\centering
	\begin{tabular}{cc}
		\includegraphics[trim={0 0.0cm 12.0cm 0cm},clip, scale = 0.55]{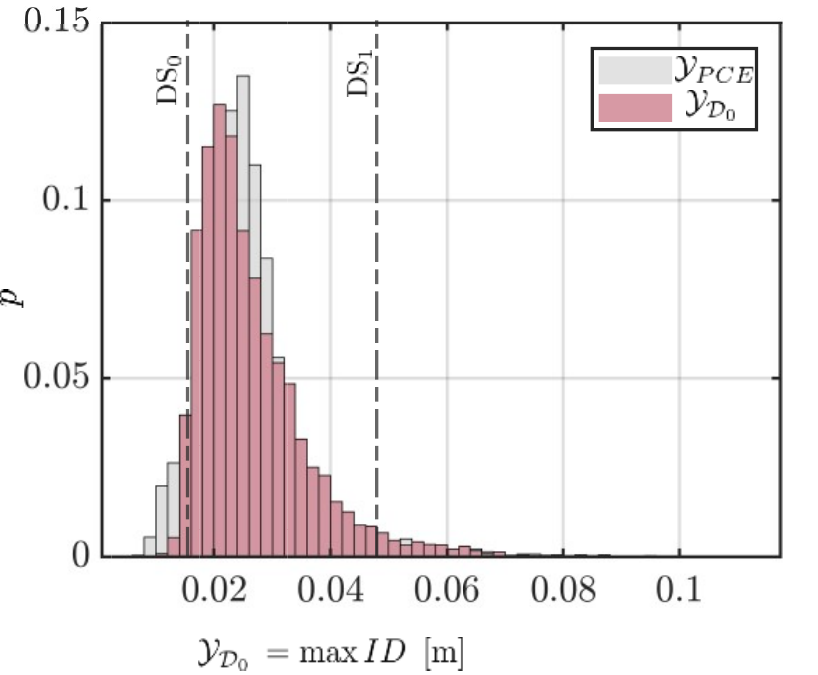} &
		\includegraphics[trim={13.5cm 0cm 0 0cm},clip, scale = 0.55]{Figures/ch3_ds0_pce.pdf} \\
		(a) &
		\hspace{15mm} (b) \\
	\end{tabular}
	\caption{(a) Comparison of the histograms for the outputs, $\mathcal{Y}_{PCE}$, of the PCE metamodel and the EDPs, $\mathcal{Y}_{\mathcal{D}_0}$, of the $\mathcal{D}_0$ dataset, i.e., the initial damage state condition $\mathrm{DS_0}$, and (b) the associated validation $\operatorname{\mathcal{Y}_{\mathcal{D}_0}-\mathcal{Y}_{PCE}}$ plot.\color{white}{ Limit state thresholds and $\pm 1 \cdot \sigma$, due to the uncertainties related to different initial damage conditions, are reported in the histograms of (a).}}
	\label{fig:ch3_PCE_ds0}
\end{figure*}

\begin{figure*}[!ht]
	\centering
	\begin{tabular}{cc}
		\includegraphics[trim={0 0.0cm 12cm 0cm},clip, scale = 0.55]{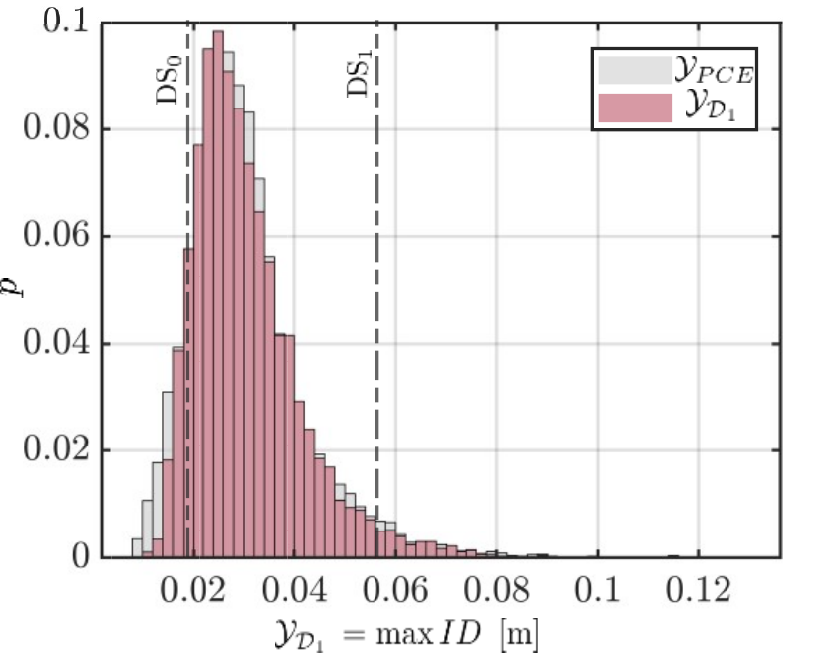} &
		\includegraphics[trim={13.5cm 0cm 0 0cm},clip, scale = 0.55]{Figures/ch3_ds1_pce.pdf} \\
		(a) &
		\hspace{15mm} (b) \\
	\end{tabular}
	\caption{(a) Comparison of the histograms for the outputs, $\mathcal{Y}_{PCE}$, of the PCE metamodel and the EDPs, $\mathcal{Y}_{\mathcal{D}_1}$, of the $\mathcal{D}_1$ dataset, i.e., the initial damage state condition $\mathrm{DS_1}$, and (b) the associated validation $\operatorname{\mathcal{Y}_{\mathcal{D}_1}-\mathcal{Y}_{PCE}}$ plot.}
	\label{fig:ch3_PCE_ds1}
\end{figure*}

\begin{figure*}[!ht]
	\centering  
	\includegraphics[width =0.99\linewidth]{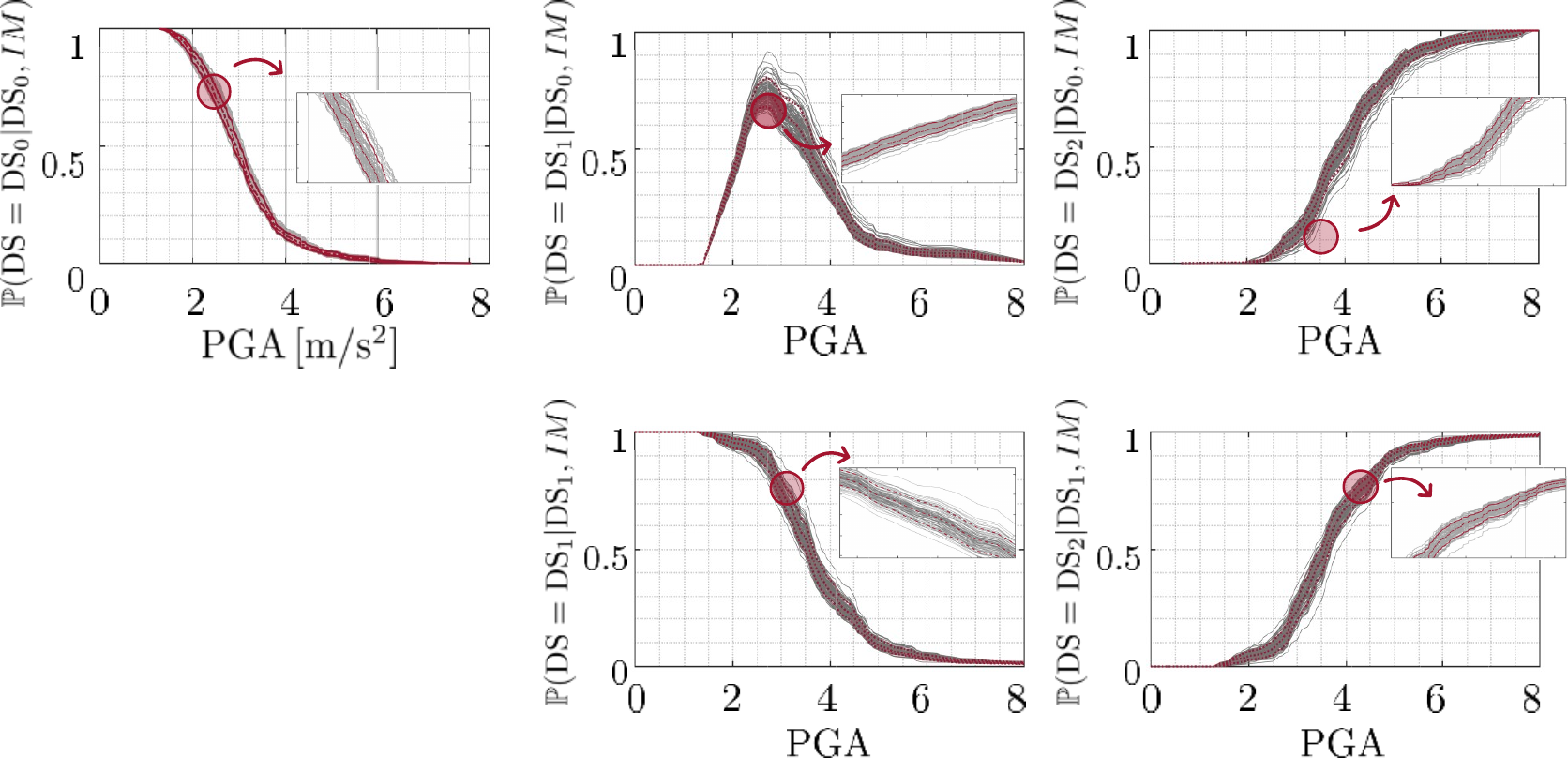}
	\caption{Bootstrap-PCE-based MCS state-dependent fragility functions: grey lines represent single seed simulations; dashed-dotted dark red lines depict the $90\%$ confidence bounds, whilst dashed-dark red ones the $50\%$.}
	\label{fig:ch3_PCE_fragility}
\end{figure*}

\noindent To test the methodology of Section~\ref{ch2:methodology}, surrogate models are developed for the MDoF model, using as input the low dimensional representation of the seismic input, that is $\widehat{\textbf{IM}}$. Specifically, PCE is adopted, since it provides a functional approximation of the computational model $\mathcal{M(\widehat{\textbf{IM}})}$ through its spectral representation on suitably built basis $\psi_{\alpha} \left( \cdot \right)$ of polynomial functions \cite{bib:UQ_pce}. In real-world problems, due to limited computing power, the bases are truncated and the governing equation becomes:
\begin{flalign}
	\mathcal{Y} \, \approx \mathcal{M}^{PCE} \left( \widehat{\textbf{IM}}\right) = \sum_{\alpha \in \mathcal{A}} c_{\alpha} \psi_{\alpha} \left( \widehat{\textbf{IM}} \right),
	\label{eq:PCE_general}
\end{flalign}
\noindent where $\mathcal{A}$ is the set of selected multi-indices of multivariate polynomials and $c_{\alpha}$ are the corresponding coefficients to be determined. The $c_{\alpha}$ coefficients can then be calculated via projection, i.e., Gaussian or sparse quadrature, or regression methods, i.e., least-square algorithms.
In this framework, we adopted the least-square method to determine the coefficients. Specifically, separated $c_{\alpha}$ were estimated for each DoE $\mathcal{D}_{s}$, constituted by the clustered pairs of pseudo-IMs and QoI of FE model response, i.e., $(\widehat{\textbf{IM}};\mathcal{Y})$, as:
\begin{flalign}
	c_{\alpha} \, = \, \text{argmin} \frac{1}{N} \sum_{i=1}^{N} \left[ y_i \, - \, \sum_{\alpha \in \mathcal{A}} c_{\alpha} \psi_{\alpha} \left( \widehat{\textbf{IM}}_i \right) \right]^{2},
	\label{eq:coeff_eqn}
\end{flalign}
where $N$ is the cardinality of the $\mathcal{D}_{s}$, $s \in [\mathrm{DS}_0;\mathrm{DS}_1]$ denotes the initial damage state level of the DoE, and $\{ y^{(1)}, \dots, y^{(N)} \}$ the realizations of $\mathcal{Y}$.
The choice of a least-square regression strategy is motivated by the possibility of adopting the bootstrap resampling method. This is particularly insightful when the size of the $\mathcal{D}_{s}$ is limited, since it allows the user to explore exhaustively the information on the variability of the dataset.
This can be achieved by first generating a set of bootstrap-resampled experimental design pairs $( \widehat{\textbf{IM}}^{(b)}, \mathcal{Y}^{(b)} )$ and then, by calculating a corresponding set of coefficients $c_{\alpha}^{(b)}$. Therefore, the response of each bootstrap PCE can be evaluated, yielding a family of $b$ surrogate models that can be interpreted as trajectories.
As recently explored in \cite{bib:marelli2018}, this process of bootstrap-based trajectories resampling can be directly employed to assess the confidence bounds on surrogate-based estimators.\\
In this case, surrogates are developed for the two sets of initial damage state conditions, i.e., $\mathrm{DS}_0$ (pristine) and $\mathrm{DS}_1$ (damaged) configurations.  As recommended in~\cite{bib:nora2022}, the subspace pursuit (SP) solver was deployed in the PCE metamodel. Nevertheless, to identify the most suitable way to calculate the coefficients $c_{\alpha}$ of the PCEs, investigations were carried out on different sizes of the $\mathcal{D}_s$ and the vector of pseudo $\widehat{\textbf{IM}}$, describing the seismic input. 
Specifically, the vector of pseudo $\widehat{\textbf{IM}}$ is defined as the $n,PCA$ most significant PCA principal components.
\begin{flalign}
	\widehat{\textbf{IM}} \, = \, \left[ \mathrm{PC}_1, \, \mathrm{PC}_2, \, \dots, \mathrm{PC}_{n,PCA} \right]^{T}.
	\label{eq:IMhat}
\end{flalign}
Figure~\ref{fig:ch2_scree_plot} reports the variability covered by increasing the number of PCs through a scree plot. As one can notice, the first 3 PCs are sufficient to describe more than the $80$\% of the input data, 6 PCs the $90$\% and 10 PCs are needed to get the $99$\% of the input variance. Since there is no notable increase in computational burden between choosing either 6 or 10 PCs, we opt for the latter. 
\FloatBarrier 
To highlight which and how each $im_i$ contributes to PCs, the first three PCs are displayed on the biplots in Figure~\ref{fig:ch2_biplot}. Particularly, in Figure~\ref{fig:ch2_biplot}, only the relevant IMs---i.e., the ones that present high scores---are plotted and labelled with the corresponding number ID of Table~\ref{tab:ch3_feature_IM}. Therefore, the information provided is twofold: (i) on the correlation among IMs; (ii) on the magnitude of the importance of the IMs in the definition of PCs.
For instance, for $PC_2$: almost all the plotted IMs are positively correlated with $PC_2$ itself. The 33-IM, i.e.,$\mathrm{F_m}$, represent the exception, since it is in the opposite direction with $PC_2$, thus entailing a negative association. The positive/negative correlations and the order of magnitude are summarized in the coordinates of the PCs next to the biplot, until the tenth coefficient.
More details on the correlations among IMs are discussed in the attached~\ref{app:A_im_feature}, along with histograms and inferred marginal distributions.
To determine the optimal size of the DoE, three measures of errors on the performances of the PCE surrogates were used: (i) the relative generalization $\varepsilon_{gen}$; (ii) the leave-one-out $\varepsilon_{LOO}$; and (iii) the empirical $\varepsilon_{emp}$ error. In particular, the former 
\begin{align}
	\displaystyle \varepsilon_{gen} \, = \, \mathbb{E} \left[ \left(\mathcal{Y}-\mathcal{M}^{PCE}\left(\widehat{\textbf{IM}}\right)\right)^2\right]/\mathrm{Var[\mathcal{Y}]}    
\end{align}
is best suited if a validation set is available; otherwise, as it is commonly the case with expensive computational models, the other two are preferred and evaluated with the following definition:
%
\begin{align}
	\varepsilon_{LOO} &= \frac{\ds \sum_{i=1}^{N_{sims}}{\left( \ve{y}_i -\mathcal{M}^{PCE\setminus i}\left(\ve{\widehat{im}}_i\right)\right)}^{2}}{\ds \sum_{i=1}^{N_{sims}} \left(\ve{y}_i-\hat{\mu}_{Y}\right)^{2}}, \\ 
	\varepsilon_{emp} \, &= \,  \frac{\ds \sum_{i=1}^{N_{sims}}{\left(\ve{y}_i-\mathcal{M}^{PCE}\left(\ve{\widehat{im}}_i\right)\right)}^{2}}{\ds \sum_{i=1}^{N_{sims}} \left(\ve{y}_i-\hat{\mu}_{Y}\right)^{2}}, 
	\label{eq:error_validation}
\end{align}
where $N_{sims}$ is the total number of PCE-based MCS, $y_i$ the realizations of $\mathcal{Y}$, and $\hat{\mu}_{Y}$ is the sample mean of the DoE response. In addition, the bootstrap technique is applied to PCE to provide a local error estimator. Precisely, resampling with substitution is used on the DoE $\mathcal{D}_s$, thus generating a set $B \, = 500$ bootstrap replications with the same number of sample points as the original $\mathcal{D}_s$. Each $b \in [1,B]$ replication is used to evaluate the corresponding PCE, yielding to $b$ different sets of the PCE coefficients and predictions. Those $b$ sets of responses are used to calculate the local variance and quantiles of the PCE predictor, exploiting all the information of the finite $\mathcal{D}_s$ size.
\begin{figure*}[!ht]
	\centering
	\begin{tabular}{cc}
 \hspace{-6.5mm}
		\includegraphics[trim = {7.5 0 17.75cm 0},clip,scale = 0.50]{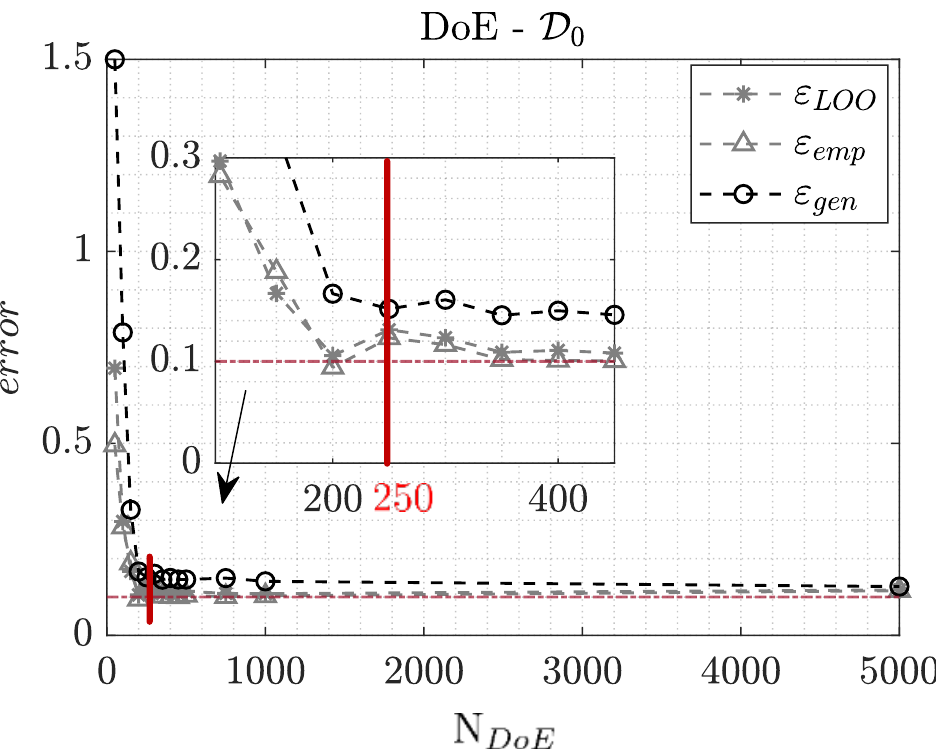} &
		\includegraphics[trim = {17.75cm 00 1. 0},clip,scale = 0.50]{Figures/ch3_MDOF_error_dataset_size_0_and_1.pdf} \\
		(a) &
		(b) \\
	\end{tabular}
	\caption{Error plots for the initial damage limit state (a) $\mathrm{DS_0}$, namely DoE $\mathcal{D}_0$, and the (b) $\mathrm{DS_1}$, namely DoE $\mathcal{D}_1$.  The error threshold is fixed to 0.10 and is represented by the red dotted horizontal line. In both datasets, the $\varepsilon_{LOO}$ and $\varepsilon_{emp}$ perform better than the $\varepsilon_{gen}$, despite the vast validation set available in DoE $\mathcal{D}_1$, see Table~\ref{tab:ch3_cluster_mdof_trx}a.}
	\label{fig:ch3_ED_error}
\end{figure*}
Figure~\ref{fig:ch3_ED_error} shows the error trends against the DoE $\mathcal{D}_s$ size. The results agree with the latest investigations reported in~\cite{bib:nora2022} concerning the ideal size of the DoE. Both the PCE-surrogate models for the $\mathcal{D}_0$ and $\mathcal{D}_1$ locate the optimal solution in 250 samples for the DoE dimension. Beyond this threshold, additional efforts to reduce model error result in marginal improvements or overfitting issues.
The estimation of the PCE coefficients for the undamaged $\mathcal{D}_0$ initial condition converged to a polynomial degree 3 with $\varepsilon_{LOO} \simeq 0.15$. Figure~\ref{fig:ch3_PCE_ds0} reports the histograms and validation $\operatorname{\mathcal{Y}_{\mathcal{D}_0}-\mathcal{Y}_{PCE}}$ plots. A validation set of $10^4$ samples is deployed, thanks to the large number of simulations carried out on the MDoF. A good agreement in terms of matching distributions between $\operatorname{\mathcal{Y}_{\mathcal{D}_0}-\mathcal{Y}_{PCE}}$ is reached; also, samples-pairs are neatly aligned on the $45^\circ$ line of the $\operatorname{\mathcal{Y}_{\mathcal{D}_0}-\mathcal{Y}_{PCE}}$ plot, indicating a favourable performance.
Similarly, for the damaged $\mathcal{D}_1$ initial condition of the system, the estimation of the PCE coefficients converged to a polynomial degree 3 with $\varepsilon_{LOO} \simeq 0.14$. Figure~\ref{fig:ch3_PCE_ds1} displays histograms of $10^4$ samples for the $\mathcal{Y}_{PCE}$ and FE $\mathcal{Y}_{\mathcal{D}_1}$ data. Again, since $\operatorname{\mathcal{Y}_{\mathcal{D}_1}-\mathcal{Y}_{PCE}}$ pairs are well aligned to the $45^\circ$ line, a good agreement is demonstrated.
Finally, state-dependent fragility curves are evaluated based on the results provided by the bootstrap PCE of both the pristine-$\mathcal{D}_0$ and the damaged-$\mathcal{D}_1$ datasets. As for the results of the brute-force MCS, fragilities are displayed referring to the optimal global IM. Thus, the $\mathrm{\beta_{eff,glob}}(IM_i)$ of Eq.~\ref{eq:global_beta} is evaluated for each transition state over the $i$ to 41 IMs. Next, the global optimal $IM^{**}$ is selected as follow $IM^{**} = \operatorname*{argmin}_{IM_i} \, \mathrm{\beta_{eff,glob}}(IM_i)$. Thus, Figure~\ref{fig:ch3_PCE_fragility} shows the state-dependent fragilities as functions of the global optimal $IM^{**}$, i.e.,
the PGA. 
\FloatBarrier
%
\FloatBarrier
\subsubsection{Validation of the framework}
\begin{figure*}[!ht]
	\centering
	\includegraphics[width=0.99\linewidth]{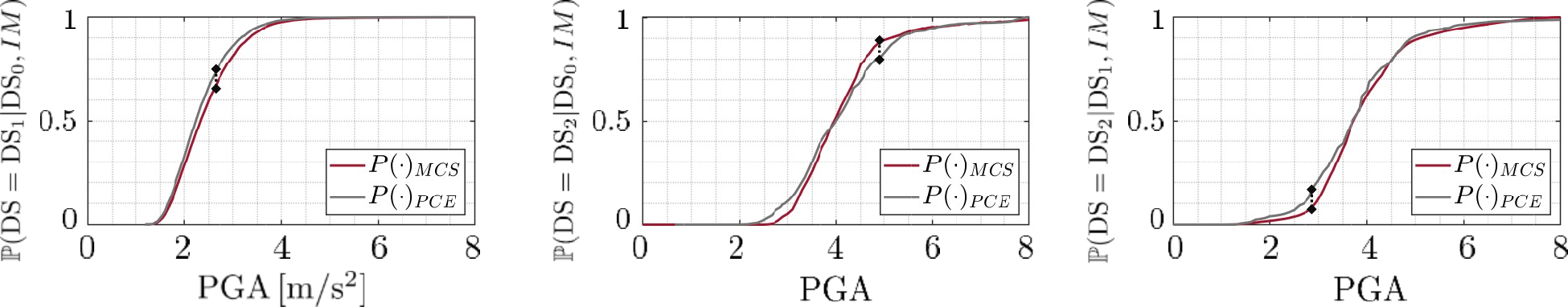}
	\caption{Mean PCE-based and brute-force MCS fragility functions comparisons for the transition states \apo0-1\apo, \apo0-2\apo, and \apo1-2\apo with PGA as IM. Note that the probability of exceedance is adopted for the transition state  \apo0-1\apo.}
	\label{fig:ch3_qualitative_comparison}
\end{figure*}

\noindent A visual comparison between state-dependent fragilities derived by brute-force MCS (benchmark case) and surrogate-based MCS shows a good agreement, see for instance Figure~\ref{fig:ch3_mcs_fragility_pga} vs Figure~\ref{fig:ch3_PCE_fragility}, respectively. 
From a quantitative point of view, we measure the difference between the two fragility functions using the efficiency indices, the Kullback–Leibler (KL) divergence, and the Kolmogorov-Smirnov (KS) distance\footnote{Technically speaking, the fragility functions are not proper CDFs, as the variable IM is in its conditional form. However, as they are assumed to be monotonically increasing, then they can be treated as CDFs, and the KL and KS distances are therefore used to assess the difference between the MCS and the PCE-based fragilities.}.  In particular, Table~\ref{tab:ch3_mdof_delta} gathers the $\Delta(\mathrm{\beta_{eff}})$ expressed as $\displaystyle \left( \vert \mathrm{\beta_{eff}^{PCE}}-\mathrm{\beta_{eff}^{}} \vert \right) /\mathrm{\beta_{eff}^{}}$ for each of the best indices of the transition states. The closer the delta is to 0, the better the matching of surrogate-based MCS with brute-force MCS results. It is generally observed that the $\Delta(\mathrm{\beta_{eff}})$ remains under the 0.20 threshold, which we consider, in this case, a good matching of responses.
\begin{table*}[!ht]
	\centering
	\caption[$\Delta(\mathrm{\beta_{eff}})$ efficiency indices for each transition states.]{$\Delta(\mathrm{\beta_{eff}})$ efficiency indices calculated as $\left( \vert \mathrm{\beta_{eff}^{PCE}}-\mathrm{\beta_{eff}^{}} \vert \right)/\mathrm{\beta_{eff}^{}}$ for the same top IMs for each transition states.}{
		\label{tab:ch3_mdof_delta}%
		\scalebox{0.65}{
			\begin{tabular}{llrllrllrllrll}
				\cmidrule{1-2}\cmidrule{4-5}\cmidrule{7-8}\cmidrule{10-11}\cmidrule{13-14}    \multicolumn{1}{c}{\textbf{Frag. \apo0-0\apo}} & \multicolumn{1}{c}{\boldmath{}\textbf{$\Delta(\mathrm{\beta_{eff}})$}\unboldmath{}} &       & \multicolumn{1}{c}{\textbf{Frag. \apo0-1\apo}} & \multicolumn{1}{c}{\boldmath{}\textbf{$\Delta(\mathrm{\beta_{eff}})$}\unboldmath{}} &       & \multicolumn{1}{c}{\textbf{Frag. \apo0-2\apo}} & \multicolumn{1}{c}{\boldmath{}\textbf{$\Delta(\mathrm{\beta_{eff}})$}\unboldmath{}} &       & \multicolumn{1}{c}{\textbf{Frag. \apo1-1\apo}} & \multicolumn{1}{c}{\boldmath{}\textbf{$\Delta(\mathrm{\beta_{eff}})$}\unboldmath{}} &       & \multicolumn{1}{c}{\textbf{Frag. \apo1-2\apo}} & \multicolumn{1}{c}{\boldmath{}\textbf{$\Delta(\mathrm{\beta_{eff}})$}\unboldmath{}} \\
				\multicolumn{1}{c}{\textbf{Optimal IM}} & \multicolumn{1}{c}{\textbf{index}} &       & \multicolumn{1}{c}{\textbf{Optimal IM}} & \multicolumn{1}{c}{\textbf{index}} &       & \multicolumn{1}{c}{\textbf{Optimal IM}} & \multicolumn{1}{c}{\textbf{index}} &       & \multicolumn{1}{c}{\textbf{Optimal IM}} & \multicolumn{1}{c}{\textbf{index}} &       & \multicolumn{1}{c}{\textbf{Optimal IM}} & \multicolumn{1}{c}{\textbf{index}} \\
				\cmidrule{1-2}\cmidrule{4-5}\cmidrule{7-8}\cmidrule{10-11}\cmidrule{13-14}    $\mathrm{PGA}$ & 0.12  &       & $\mathrm{I_{RG,a}}$ & 0.13  &       & $\mathrm{ASI}$ & 0.14  &       & $\mathrm{ASI}$ & -0.17  &       & $\mathrm{E-ASA_{R80}}$ & 0.07 \\
				$\mathrm{IC}$ & 0.15  &       & $\mathrm{PGA}$ & 0.14  &       & $\mathrm{EPA}$ & 0.13  &       & $\mathrm{EPA}$ & -0.19  &       & $\operatorname{\mathrm{E-ASA_{R200}}}$ & 0.08 \\
				$\mathrm{I_{RG,a}}$ & 0.15  &       & $\mathrm{I_{RG,v}}$ & 0.16  &       & $\mathrm{RMS{(\ddot{u})}}$ & 0.15  &       & $\mathrm{PGA}$ & -0.19  &       & $\operatorname{\mathrm{E-ASA_{R100}}}$ & 0.08 \\
				$\operatorname{\mathrm{E-ASA_{R200}}}$ & 0.12  &       & $\operatorname{\mathrm{E-ASA_{R200}}}$ & 0.16  &       & $\mathrm{PGA}$ & 0.14  &       & $\mathrm{I_{RG,v}}$ & -0.17 &       & $\operatorname{\mathrm{E-ASA_{R150}}}$ & -0.08 \\
				$\mathrm{RMS{(\ddot{u})}}$ & -0.05 &       & $\mathrm{PGD}$ & 0.16  &       & $\operatorname{\mathrm{E-ASA_{R200}}}$ & 0.15  &       & $\mathrm{I_{RG,a}}$ & -0.15 &       & $\mathrm{I_{RG,a}}$ & -0.16 \\
				\cmidrule{1-2}\cmidrule{4-5}\cmidrule{7-8}\cmidrule{10-11}\cmidrule{13-14}    \end{tabular}%
	}  }
\end{table*}
The KL divergence, $D_{\text{KL}}(P \parallel Q)$, quantifies how one probability distribution $P$ is different from a second $Q$, defined on the same sample space. Specifically, $P$ denotes the empirical probability distribution (Figure~\ref{fig:ch3_PCE_fragility}); while $Q$ represents the empirical probability distribution of the mean brute-force MCS state-dependent fragillities (Figure~\ref{fig:ch3_mcs_fragility_pga}). Thus, KL is evaluated as provided in Eq.~\ref{eq:KL}: 
\begin{align}
	D_{\text{KL}}(P \parallel Q) = \sum_{ x \in \mathcal{X} } P(x)\ \log\left(\frac{\ P(x)\ }{ Q(x) }\right)
	\label{eq:KL}
\end{align}
where $\mathcal{X}$ represents the sample space, that is, the domain of the optimal IM for each transition state. Figure~\ref{fig:ch3_qualitative_comparison} shows the mean PCE-based and brute-force MCS state-dependent fragility functions for the transition states \apo0-1\apo, \apo0-2\apo, and \apo1-2\apo. As Table~\ref{tab:KL-KS} shows, the $D_{\text{KL}}$ distance is bounded $\in [0.12-0.19]$, meaning an overall good agreement (see \cite{bib:perez2008}).
The KS distance is defined as
\begin{align}
	D_{\text{KS}} =\sup_x |F_{1}(x)-F_{2}(x)|,
	\label{eq:KS}
\end{align}
where $F_{1}$ and $F_{2}$ are the mean empirical cumulative distribution functions of the MCS brute-force and surrogate-based fragilities, respectively, and $\sup$ is the supremum function. Thus, Table~\ref{tab:KL-KS} collects the results. The distance between the mean of the MCS brute-force functions and the surrogate-based is considered acceptable within the simulation framework. Moreover, fragilities must be integrated with the hazard curves; therefore, we consider these minimal differences as negligible. 
\begin{table}[htbp]
	\centering
	\caption{$D_{\text{KS}}$ and $D_{\text{KL}}$ measures for the mean PCE-based \textit{vs} brute-force MCS state-dependent fragility functions.}
	\label{tab:KL-KS}%
	\scalebox{0.90}{
		\begin{tabular}{lccc}
			\toprule
			&        \multicolumn{3}{c}{\boldmath{}\textbf{IM =  $\mathrm{PGA}$}\unboldmath{}} \\
			& \textbf{\apo0-1\apo} & \textbf{\apo0-2\apo} & \textbf{\apo1-2\apo} \\
			\midrule
			$D_{\text{KS}}$  & 0.067 & 0.085  & 0.096 \\
			$D_{\text{KL}}$  & 0.164  & 0.200 & 0.186 \\
			\bottomrule
	\end{tabular}}
\end{table}%
\section{Industrial case: vertical tank installed on a 3D braced-frame structure} \label{ch4:application}
The above framework was applied to the SPIF \#2 case study, \cite{bib:quinci2023},\cite{bib:nardin2022},\cite{bib:butenweg-SPIF}. Specifically, the object of the state-dependent fragility analysis was the vertical tank installed on the steel-BF industrial substructure. 

\subsection{Step $\mathcal{A}$ - Computational model description}\label{sec:SPIF}
\subsubsection{FE physics-based model}
\begin{figure*}[!ht]
	\centering
	\includegraphics[scale = 0.8]{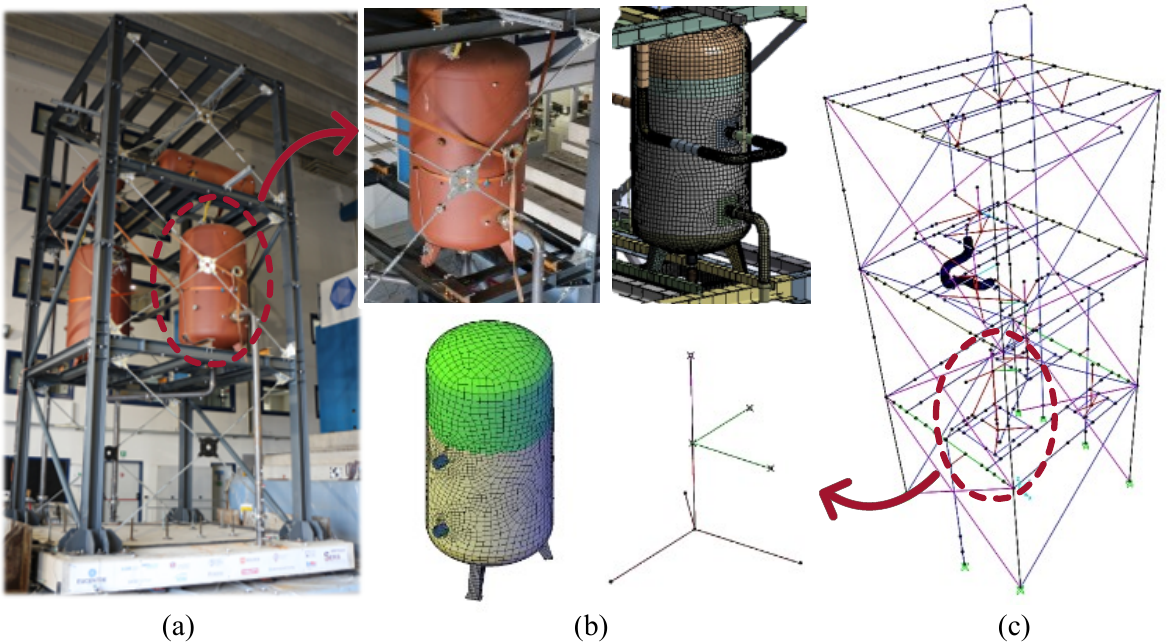}
	\caption{The SPIF \#2 mock-up: (a) photo of the braced frame (BF) configuration on the shake table of EUCENTRE Facilities and details of the vertical tank installed at the first level; (b) SAP2000\textregistered high-fidelity local FE model and \textit{ad-hoc} implemented stick-model for the vertical tank; (c) global SAP2000\textregistered FE model of SPIF \#2.}
	\label{fig:ch4_fe_shake_3}
\end{figure*}

\noindent We consider the FE model developed for the SPIF \#2 experimental research campaign. Figure~\ref{fig:ch4_fe_shake_3}(a) shows the industrial substructure model tested on the shaking table. The mock-up consists of a full-scale 3-storey steel BF structure with flexible floors. Several NSCs of the industrial process were installed, like tanks, cabinets, bolted flange joints, and T-joints. Particularly, the complex dynamic interaction between the main steel structure and one of the NSCs---the vertical tanks mounted on the first level---was examined. The strong displacement exhibited by those components deserved attention. This is because their damage can lead to severe consequences. 
Figure~\ref{fig:ch4_fe_shake_3}(a) shows the investigated vertical tank. Figure~\ref{fig:ch4_fe_shake_3}(b) reports a detail of the \textit{ad-hoc} designed stick model for the tanks and its expensive reference high-fidelity local FE model. The stick model was developed with the goal to: (i) mimic the modal properties and (ii) catch the different effects of the participating seismic mass on the supporting girders system. The latter is particularly delicate since the participating mass varies with the intensity level of the seismic excitation. A thorough discussion on FE modelling and calibration of the global SPIF \#2 model of Figure~\ref{fig:ch4_fe_shake_3}(c) is reported in~\cite{bib:quinci2023}. 
The shake table tests showed that the vertical tanks were one of the most critical elements among the installed components. Therefore, we developed fragility analysis for these elements, using the bootstrap PCE technique of the previous Section.

\subsection{Step $\mathcal{B}$ - Input definition}
The framework outlined in Section~\ref{ch2:methodology} is adopted. Sequences of seismic events were assigned to the FE model of SPIF \#2 to (i) simulate different damage initial configurations and (ii) cover each transition state. A total of 100 gms sequences, consisting of chains of 5 earthquakes each, were implemented by the same ensemble of gms generated by the GMM of Subsection~\ref{sec:input-gmm}. The computational cost pro single sequential NLTHA is $\sim$30 min on an Intel(R) Core(TM) i9-10900K CPU @3.70GHz, 10 Core(s) - 128 GB RAM.
Both the number of gms in a single sequence and the total number of sequential NLTHAs were determined by a trade-off between the total time required for analysis and a well-defined population of the transition matrix of Figure~\ref{fig:transitionStateGraph}(a).
The transition matrix for the vertical tank was defined on two performance limit states, e.g., $\mathrm{DS_1}$ and $\mathrm{DS_2}$. Those were set according to literature recommendations, see~\cite{bib:vathi2017}, and confirmed by the experimental campaign, see~\cite{bib:quinci2023}.  
The first, $\mathrm{DS_1}$, is the design basis earthquake (DBE), which is linked to the operation and functionality of the process plant. The second, $\mathrm{DS_2}$, is the safe shutdown earthquake (SSE) limit state, for which the fundamental safety functions can be ensured with minor damages, although the facility is no longer operational. Thresholds were experimentally identified by maximum acceleration values recorded at the base of the vertical tank 
\begin{equation} \label{eq:threshold_tank} 
	\mathrm{max} \, \displaystyle \vert \ddot{u}_{base}^{v.tank} \vert = \begin{cases}
		10 \, \mathrm{m/s^2}, & \text{DBE}  \, \mathrm{(DS_1)}   \\ 
		16 \, \mathrm{m/s^2}, & \text{SSE}  \, \mathrm{(DS_2)} 
	\end{cases}
\end{equation}
for the DBE and SSE, respectively, as reported in ~\cite{bib:quinci2023}.
\subsection{Step $\mathcal{C}$ - QoI response}
\subsubsection{NLTHAs and experimental design generation}
\begin{figure*}[!ht]
	\centering
	\begin{tabular}{cc}
		\includegraphics[scale = 0.45]{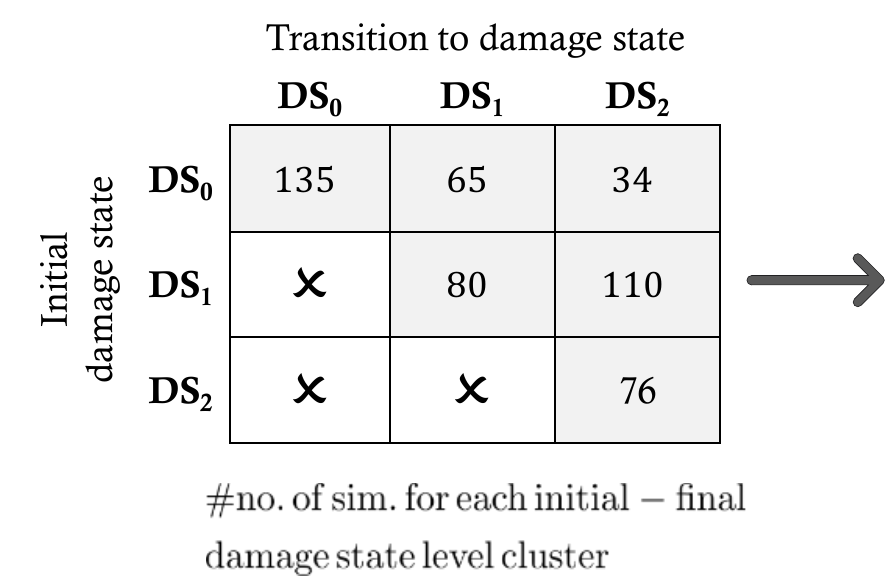} &
		\includegraphics[scale = 0.44]{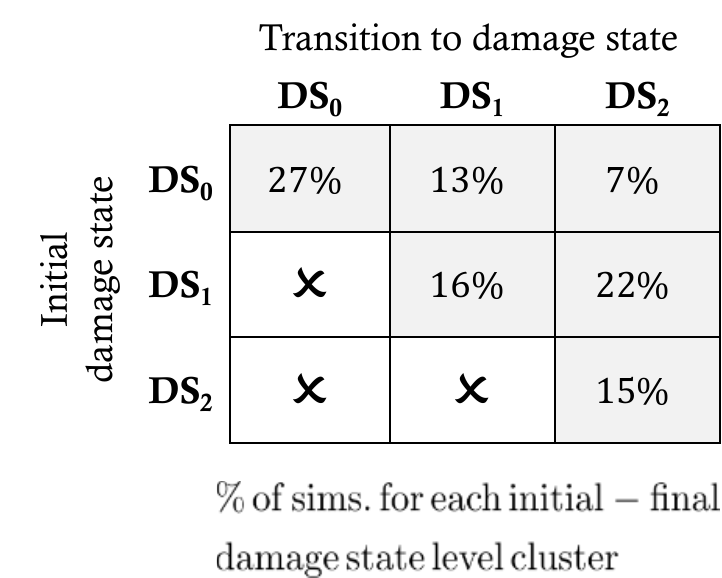} \\
		\hspace{05mm} (a) &
		\hspace{15mm} (b) \\
	\end{tabular}
	\caption{Transition state matrices for the SPIF \#2: (a) counters and (b) percentages of simulation for each initial-final damage state level cluster, respectively.}
	\label{tab:ch4_cluster_SPIF_trx}
\end{figure*}

\noindent As illustrated in Figure~\ref{fig:heuristics}, the results of the NLTHAs performed on the FE model were clustered into the transition states of Figure~\ref{fig:transitionStateGraph}(a). Specifically, $\mathcal{D}_0$ indicates the dataset associated with pristine initial damage conditions, whilst $\mathcal{D}_1$ denotes the dataset for which the threshold associated with the DBE limit state was exceeded. Lastly, $\mathcal{D}_2$ collects the results of simulations for which the SSE limit state was attained. 
As Table~\ref{tab:ch4_cluster_SPIF_trx} shows, almost 47\% of the simulations belong to the $\mathcal{D}_0$ damage state initial condition. Particularly, 13\% and 7\% transitioned from undamaged initial conditions to exceeding the DBE and SSE threshold at the end of the NLTHAs, respectively. Instead, 38\% of simulations belong to $\mathcal{D}_1$, out of which 16\% stayed in the same damage level even after other seismic shocks. Finally, 15\% of the simulations reached the absorption state. 
Again, the 41 IMs of Table~\ref{tab:ch3_feature_IM} were evaluated for each gms of the NLTHA simulations. PCA was then applied to reduce the order of the input dimension. Specifically, 10 PCs were used to cover 99\% of the variability of data.
\subsubsection{PCE metamodelling}
\begin{figure*}[!ht]
	\centering
	\begin{tabular}{cc}
		\includegraphics[trim={0cm 0cm 12.4cm 0},clip,scale = 0.450]{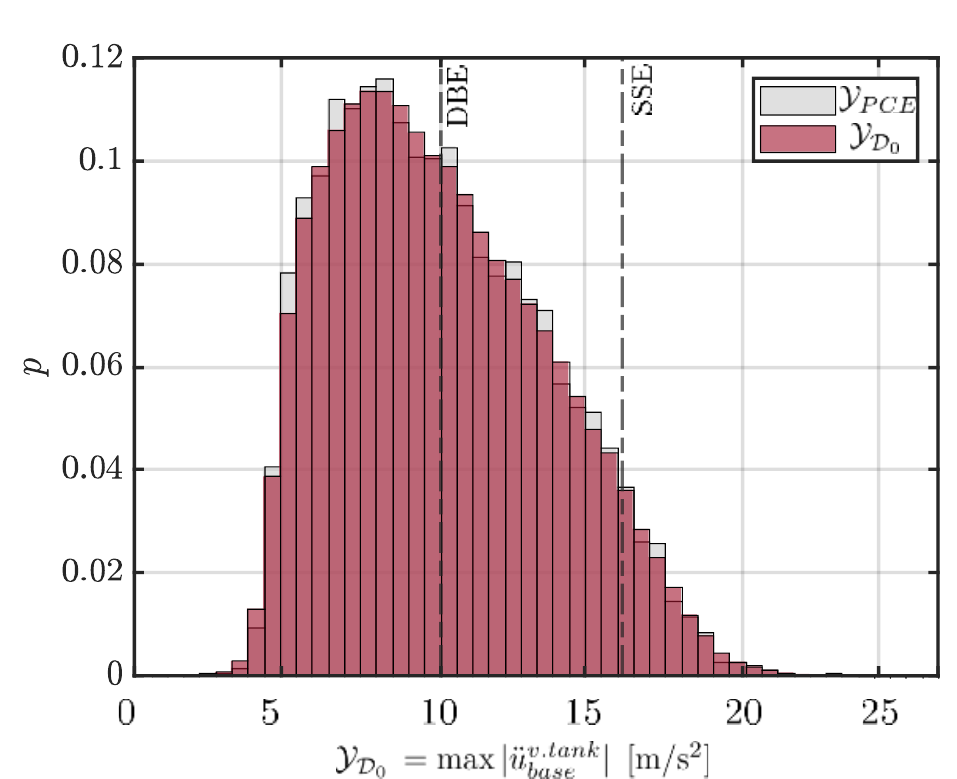} &
		\includegraphics[trim={16cm 0cm 0cm 0},clip,scale = 0.450]{Figures/ch4_spif_ds_0.pdf} \\
		(a) &
		(b) \\
	\end{tabular}
	\caption{(a) Histogram distribution of the PCE surrogate $\mathcal{Y}_{PCE}$ (light grey) predictor \textit{vs} the $\mathcal{Y}_{\mathcal{D}_{0}}$ original data of the initial undamaged condition dataset; (b) control plot of the performance of the $\mathcal{Y}_{PCE}$ surrogate model \textit{vs} the $\mathcal{Y}_{\mathcal{D}_{0}}$ reference experimental design samples.}
	\label{fig:ch4_init0}
\end{figure*}

\begin{figure*}[!ht]
	\centering
	\begin{tabular}{cc}
		\includegraphics[trim={0cm 0cm 12.25cm 0},clip,scale = 0.450]{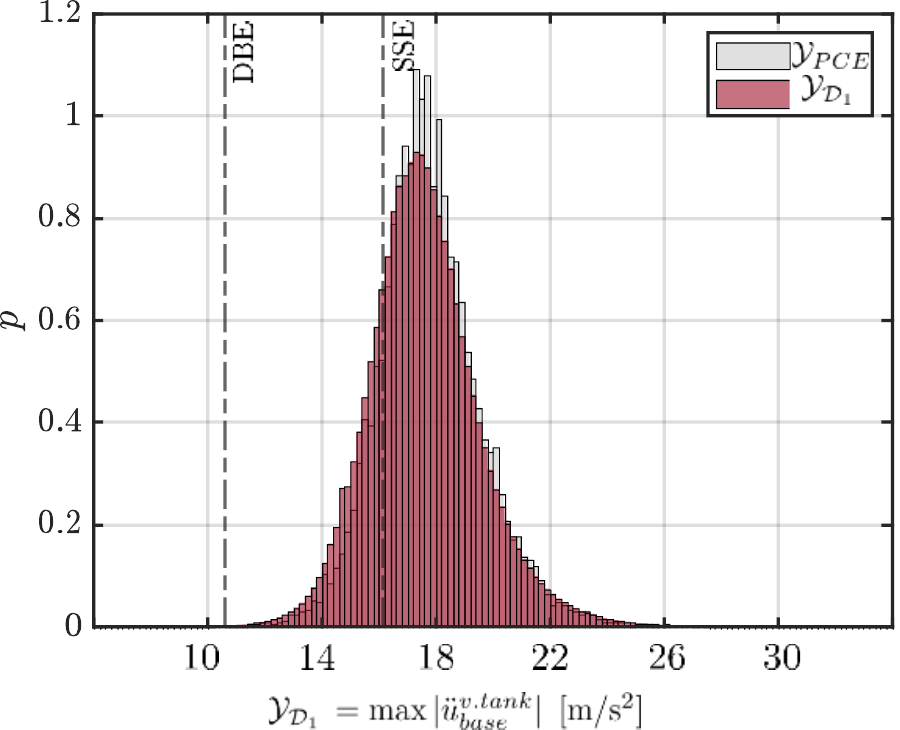} &
		\hspace{5mm}\includegraphics[trim={15.25cm 0cm 0cm 0},clip,scale = 0.450]{Figures/ch4_spif_ds_1.pdf} \\
		(a) &
		(b) \\
	\end{tabular}
	\caption{(a) Histogram distribution of the PCE surrogate $\mathcal{Y}_{PCE}$ (light grey) predictor \textit{vs} the $\mathcal{Y}_{\mathcal{D}_{1}}$ original data of the initial undamaged condition dataset; (b) control plot of the performance of the $\mathcal{Y}_{PCE}$ surrogate model \textit{vs} the $\mathcal{Y}_{\mathcal{D}_{1}}$ reference experimental design samples.}
	\label{fig:ch4_init1}
\end{figure*}

\noindent State-dependent fragilities required many more analyses than the ones performed on the expensive-to-run FE model. Hence, two metamodels were implemented to overcome the computational and time issues. Specifically, the first metamodel was built on the experimental design $\mathcal{D}_{0}$ of simulations belonging to the pristine initial condition dataset. The second on the DoE $\mathcal{D}_{1}$, the family of simulations with initial conditions attaining the DBE limit state. Resampling with substitution was used on both the experimental design $\mathcal{D}_{0,1}$, thus generating sets of bootstrap replications to estimate confidence bounds.
Specifically, bootstrap PCE was carried out with a q-norm truncation of $q = 0.50$, maximum allowed interaction $r = 2$ and a total number of replications $B = 500$. Based on the latest findings of \cite{bib:nora2022}, the subspace pursuit (SP) solver was adopted with 10 PCs.
As a result, Figure~\ref{fig:ch4_init0}(a) reports the histogram of the surrogate predictor over the histogram distribution of the $\mathcal{D}_{0}$. The limit state thresholds of DBE and SSE are also reported. A favourable performance is attained, with a final $\varepsilon_{LOO}$ error of $4.17$E-02. The estimation of bootstrap PCE coefficients converged to a polynomial degree order of 3. Moreover, Figure~\ref{fig:ch4_init0}(b) shows the control $\mathcal{Y}_{\mathcal{D}_{0}}$-$\mathcal{Y}_{PCE}$ plot: a good alignment is found with the $45^\circ$ line, which represents the ideally perfect match between true and surrogate data.
Similarly, bootstrap PCE was performed for the experimental design $\mathcal{D}_{1}$. Anew, a q-norm truncation $q = 0.50$, maximum allowed the interaction $r = 2$ and a number of replications $B = 500$ was applied. 
As a result, Figure~\ref{fig:ch4_init1}(a) reports the histogram distribution of both the $\mathcal{Y}_{PCE}$ predictor and the $\mathcal{Y}_{\mathcal{D}_{1}}$ data of the damaged initial state condition. A generally good agreement is found, with a few exceptions in the range between $[21-22] m/s^2$. This is reflected also in the control plot $\mathcal{Y}_{PCE}$-$\mathcal{Y}_{\mathcal{D}_{1}}$ of Figure~\ref{fig:ch4_init1}(b), where the  dispersion is greater in that range. The estimation of the PCE coefficients converged at a polynomial degree of order 5 with a final $\varepsilon_{LOO}$ error of $4.73$E-02.\\
\FloatBarrier
\subsection{Fragility assessment}
\begin{figure*}[!ht]
	\centering
	\includegraphics[width = 0.95\linewidth]{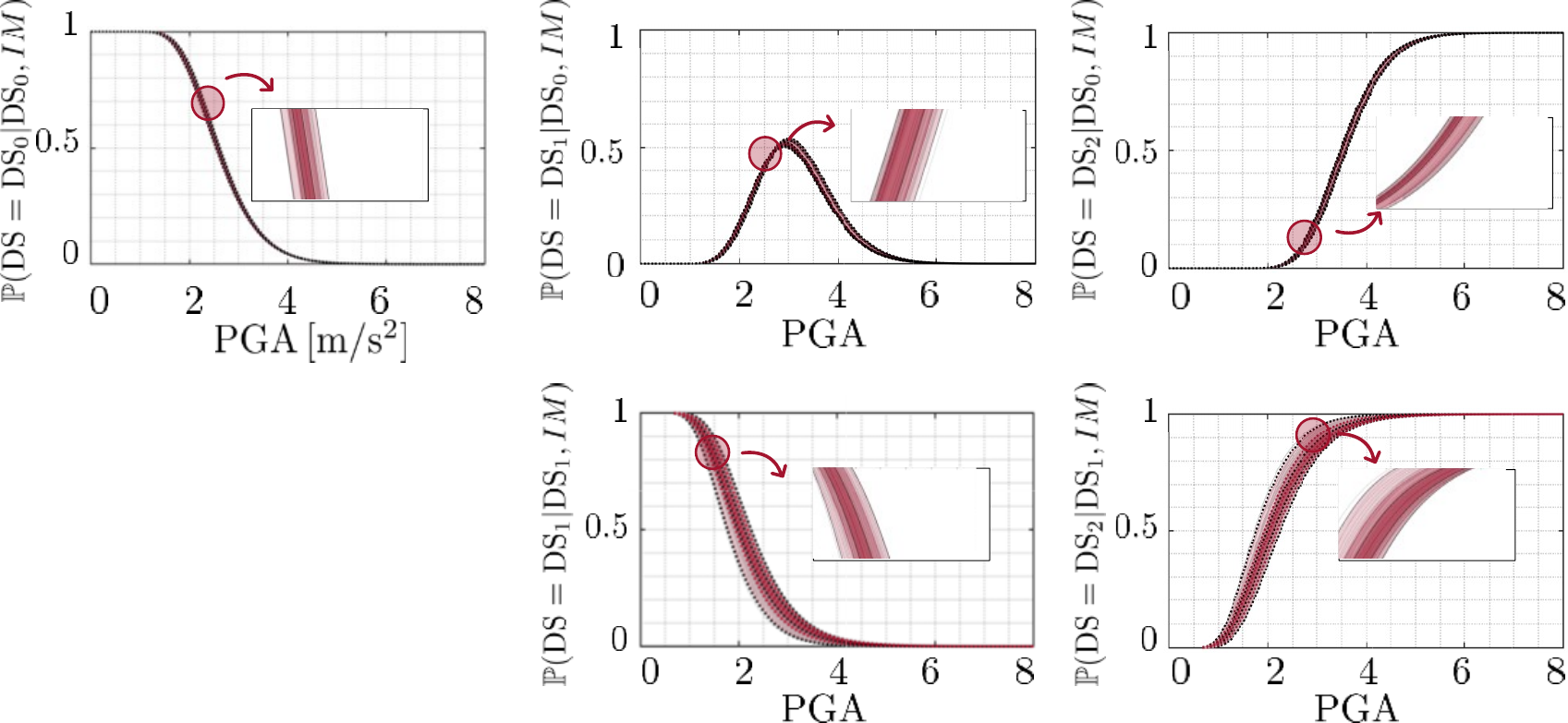}
	\caption{Bootstrap-PCE state-dependent fragility curves of the SPIF \#2 vertical tank: black-dotted thick lines stand for the 1\%, 50\%, and 99\% percentiles. Red-dotted lines for the 10\% and 90\% percentiles, along with their area with darker to lighter red-shaded colours.}
	\label{fig:ch4_fragility_PGA}
\end{figure*}

Following Section~\ref{ch3:benchmark}, a global optimal IM descriptor for the derivation of fragility functions is evaluated. Table~\ref{tab:ch4_global_beta_SPIF} gathers the ranking of the first five minimum $\beta_{\text{eff,glob}}$ for the SPIF case study, derived by Eq.~\ref{eq:global_beta}.
It emerges that IMs strictly correlated with acceleration or energy content are the most suitable for the case study. Next, we select PGA as the optimal IM and compute state-dependent fragilities using PCE-surrogate models as described in Section~\ref{ch2:methodology}.
\begin{table}[htbp]
	\centering
	\scalebox{0.75}{
		\begin{tabular}{llc}
			\toprule
			& \multicolumn{1}{c}{\textbf{Global }} & \boldmath{}\textbf{$\beta_{\text{eff,glob}}$}\unboldmath{} \\
			& \multicolumn{1}{c}{\textbf{Optimal IM}} & \textbf{index [\%]} \\
			\midrule
			1     & $\mathrm{PGA}$ & 4.02 \\
			2     & $\mathrm{E_{cum}}$ & 4.35 \\
			3     & $\mathrm{I_{A}}$ & 5.07 \\
			4     & $\mathrm{I_{F}}$ & 5.31 \\
			5     & $\mathrm{I_{RG,a}}$ & 7.42 \\
			\bottomrule
	\end{tabular}}
	\captionof{table}{Ranking of the optimal IMs according to the global $\mathrm{\beta_{\text{eff,glob}}}$, defined in Eqn.~\ref{eq:global_beta}.} 
	\label{tab:ch4_global_beta_SPIF}    
\end{table}%
\FloatBarrier
Figure~\ref{fig:ch4_fragility_PGA} shows the probabilistic description of state-dependent fragilities as functions of the PGA. The curves representing the 1\%-50\%-99\% percentiles are displayed with black-dotted thick lines, whilst the 10\%-90\% with red-dotted lines, and their area with darker to lighter red-shaded colours.
In \ref{app:D_SPIF_collection}, we report the collection of state-dependent fragility functions based on the optimal $\beta_{\text{eff}}$ index for each transition state in Figure~\ref{fig:appC_PCE_fragility}. Moreover, the commonly referred fragility functions as the probability of exceedance of a certain threshold, in this case, the $\mathrm{DS_1}$ threshold given $\mathrm{DS_0}$ initial damage condition, is reported in 
Figure~\ref{fig:D_SPIF_01_PGA}.
\FloatBarrier
%
\section{Conclusions and Future developments}\label{ch5:conclusion}
An innovative, non-intrusive UQ-based framework for assessing state-dependent fragility functions has been proposed. The framework builds on a stochastic representation of seismic sequences, calibrated and validated FE models, principal component analysis (PCA) representation of the input, and advanced polynomial chaos expansion (PCE)-based surrogate modelling. The proposed framework provides for both vast flexibility and integration with FEM experts as well as extreme computational efficiency. 
State-dependent fragility analysis requires a vast number of NLTHAs, based on time series of stochastic seismic sequences. However, when considering realistic computational models, the significant computing demands of an extensive set of sequential NLTHAs generally hinder this direct derivation of state-dependent fragilities. Hence, we propose an innovative UQ-based framework that combines a PCA representation of the stochastic input and an advanced PCE surrogate modelling of the QoI of the system. Specifically, we first performed a reduced number of stochastic seismic sequence NLTHAs on expensive-to-run FEM. Next, the resulting QoIs were clustered according to predefined damage initial states. At the same time, we performed PCA on the exhaustive vector of $\boldsymbol{IM}$, to obtain a low-dimensional input representation of the time series sequences $\widehat{\boldsymbol{IM}}$. Next, on the pairs of clustered QoIs and the $\widehat{\boldsymbol{IM}}$---i.e., the DoE---, we built different PCE surrogate models, one for each initial damage state of the system. Successively, the vast number of MCS surrogate-based analyses enabled us to derive non-parametric state-dependent fragility functions. 
In particular, the UQ-based framework was applied twice. First, it was tested and validated on a simple yet realistic 2D MDoF system endowed with Bouc-Wen hysteresis. Specifically, given the inexpensive-to-run benchmark case study, state-dependent fragilities were evaluated both via the MCS brute-force method and the MCS PCE-based one. Global and local efficiency $\mathrm{\beta_{eff}}$ indeces for each transition states were evaluated to determine the optimal IM for fragility assessment. Moreover, qualitative and quantitative comparisons through $\mathrm{\Delta\beta_{eff}}$ and statistical measures confirmed the acceptable performance of the developed framework. 
Second, the validated methodology was applied to derive seismic state-dependent fragility functions for an industrial process component. Specifically, the critical vertical tank of the 3-storey 3D BF industrial mock-up of project SPIF \#2 was considered. Following the previous example, a given number of sequences of synthetic ground motions were assigned as input for NLTHAs on the refined FEM of the coupled SPIF system. Then, clustering of the QoIs along with PCA for dimensionality reduction of the stochastic seismic sequences of the input was performed. Next, PCE surrogate models were built for the identified initial damage state conditions. Moreover, MCS PCE-based state-dependent fragility functions were evaluated. 
Thus, the developed framework allows to unlock the possibility of efficiently computing state-dependent fragility for a variety of problems. In addition, the versatility of the framework allows us to extend it to a vector of IMs for fragility assessment, in the future. The use of the framework for aftershock sequences is being considered as a second future direction, provided with a stochastic representation of input sequences. Finally, either the single state-dependent fragility function or the entire framework can be used to estimate seismic risk.
\section*{Acknowledgements}
{The research leading to these results has received funding from (i) the European Community's “NextGeneration EU Programme” [$\mathrm{PRIN-2022MJ82MC\_001 - CUP E53D23004440006}$] - for the first and the last author; (ii) the National Project MUR PNRR M4C2-CN1-SPOKE 9 - for the first and the third author; and (iii) the Italian Ministry of Education, University and Research (MIUR) in the frame of the “Departments of Excellence 2023-2027” (grant $L232/2016$) - for the third and the last author.}
%
\newpage
\onecolumn
\appendix
\section{- Histograms and distributions of the input.} \label{app:A_im_feature}
\setcounter{figure}{0}                       
\renewcommand\thefigure{A.\arabic{figure}}   
Figure~\ref{fig:heatmap_IM} illustrates the correlation among the IM parameters, which varies $\left[ -1, +1 \right]$. It emerges:
\begin{itemize}
	\item high correlations among $\operatorname{\mathrm{E-ASA_{R,x}}}$, $\mathrm{Sa}$ and $\operatorname{\mathrm{E-ASA_{R,x}}}$ as the frequency drop $R_x$ increases - indexes from 36 to 41; 
	\item significative correlations among IMs sensitive to acceleration $\mathrm{EPA}$ - idx. 30 - , $\mathrm{ASA_{40}}$ - idx. 28, and $\mathrm{ASI}$ - idx. 29;
	\item negative correlations or low correlations for the significant time duration $\mathrm{T_d}$ - idx. 23, the Cosenza-Manfredi intensity $\mathrm{I_{CM}}$ - idx. 27 - and both the frequency-related IMs mean $\mathrm{F_m}$ - idx. 33 - and rate of change mean frequencies $\mathrm{\dot{F}_m}$ - idx. 34;
	\item linear correlations between $\mathrm{Sa}$-$\mathrm{Sv}$ and $\mathrm{Sv}$-$\mathrm{Sd}$ - idxs. 4-18, as expected by the definitions.
\end{itemize}
\begin{figure*}[!ht]
	\centering
	{\includegraphics[trim = 0cm 0 0.0cm 0cm,clip,scale = 0.25]{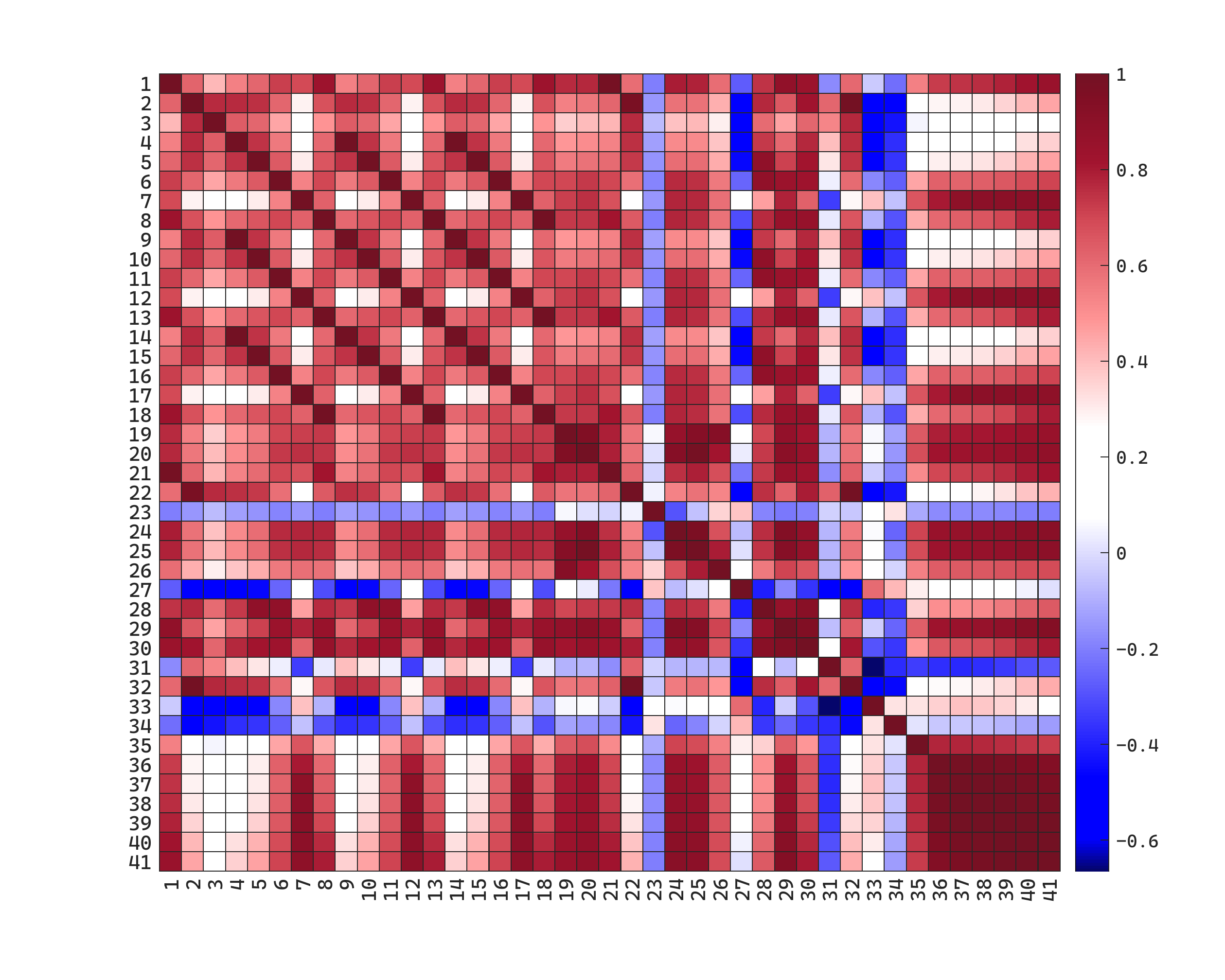}}
	\caption{Correlation map between input IM parameters.}
	\label{fig:heatmap_IM}
\end{figure*}
Table~\ref{tab:app_A_IM_distribution} and Figures~\ref{fig:IM_distr_1_16}-\ref{fig:IM_distr_37_41} gather the histograms and inferred \pdf~ for each IM presented in Table~\ref{tab:ch3_feature_IM} of Section~\ref{sec:input-gmm}. 
\input{Tables/app_A_IM_distribution}
\begin{figure*}[!ht]
	\centering
	{\includegraphics[width=0.9\textwidth]{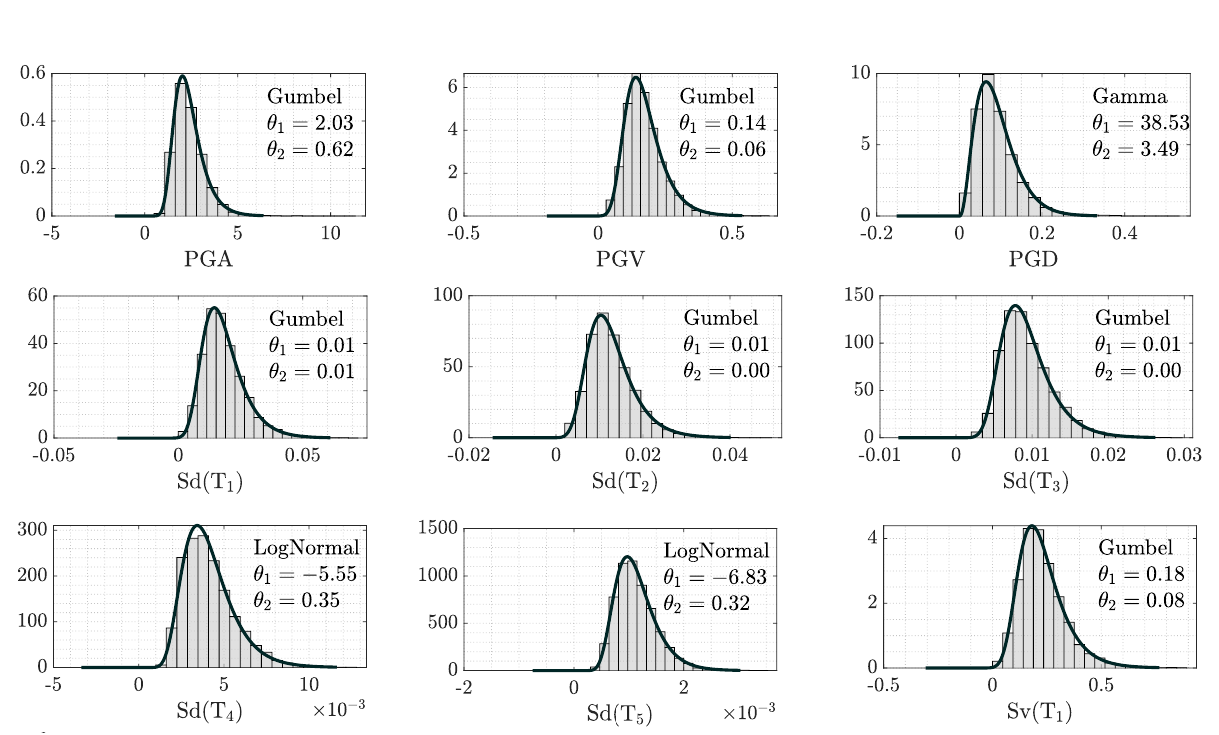}}
	\caption{Histograms and distributional models for IMs from 1 to 9.}
	\label{fig:IM_distr_1_16}
\end{figure*}
\begin{figure*}[!ht]
	\centering
	{\includegraphics[width=0.9\textwidth]{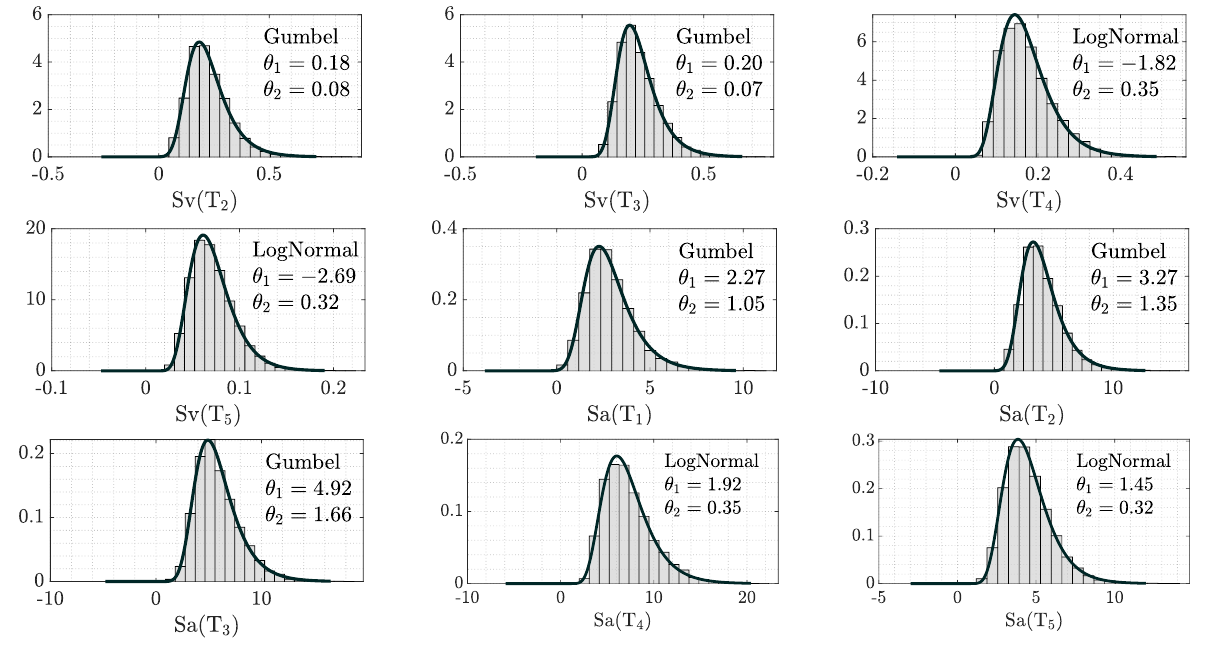}}
	\caption{Histograms and distributional models for IMs from 10 to 18.}
	\label{fig:IM_distr_10_18}
\end{figure*}
\begin{figure*}[!ht]
	\centering
	{\includegraphics[width=0.9\textwidth]{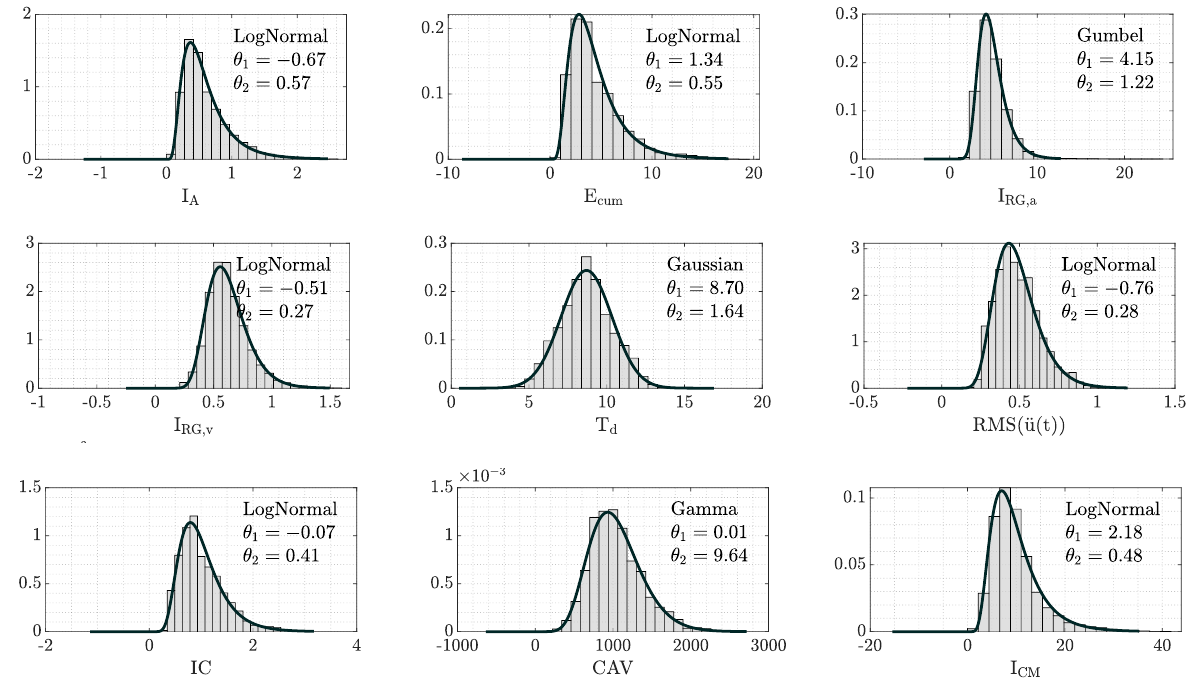}}
	\caption{Histograms and distributional models for IMs from 19 to 27.}
	\label{fig:IM_distr_19_27}
\end{figure*}
\begin{figure*}[!ht]
	\centering
	{\includegraphics[width=0.9\textwidth]{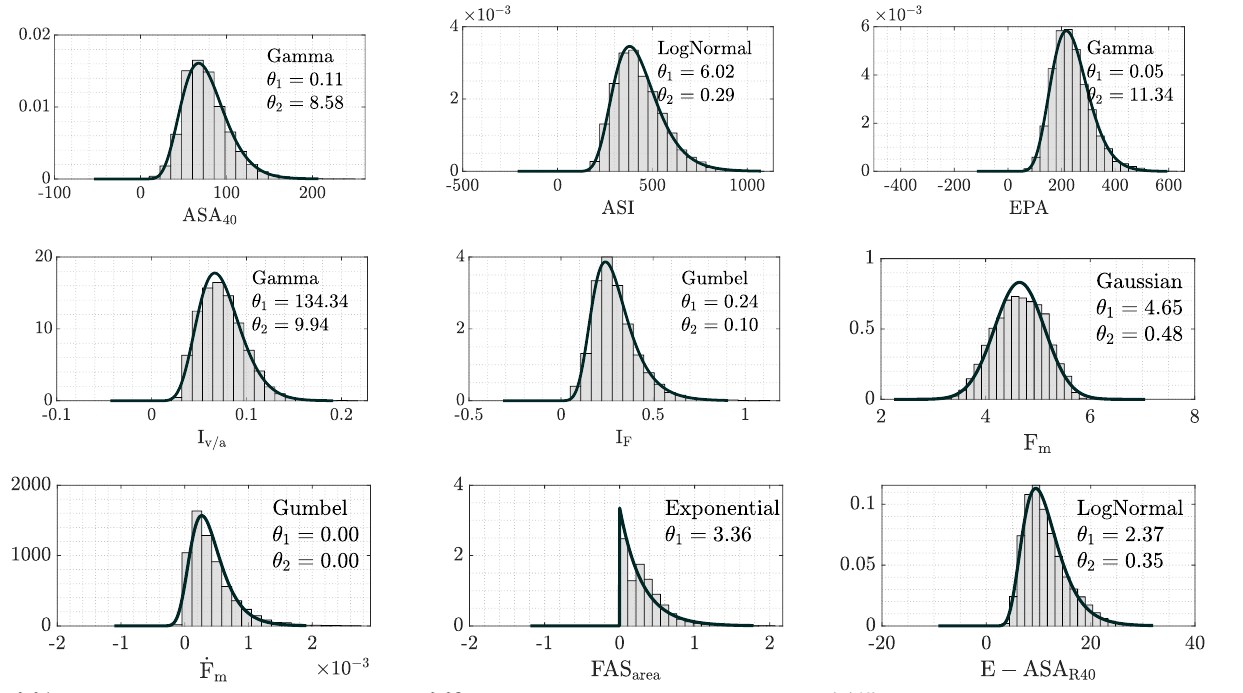}}
	\caption{Histograms and distributional models for IMs from 28 to 36.}
	\label{fig:IM_distr_28_36}
\end{figure*}
\begin{figure*}[t!]
	\centering
	{\includegraphics[width=0.9\textwidth]{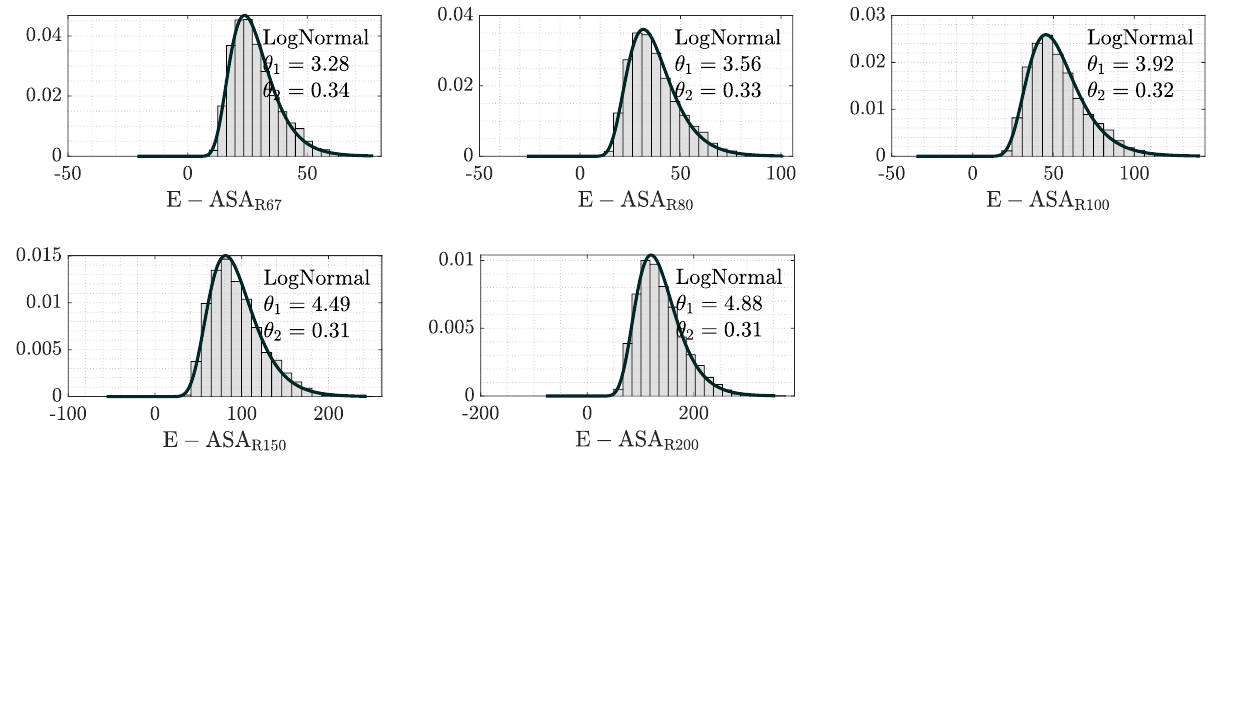}}
	\caption{Histograms and distributional models for IMs from 37 to 41.}
	\label{fig:IM_distr_37_41}
\end{figure*}
\FloatBarrier

\section{- MDoF support material} \label{app:B_MDOF_collection}
%
\setcounter{figure}{0}                       
\renewcommand\thefigure{B.\arabic{figure}}   
\subsection*{State-dependent fragility: $\mathbb{P}(\mathrm{DS}\geq \mathrm{DS_1}|\mathrm{DS_0}, IM)$.} 

\begin{figure}[!ht]
	\centering
	\includegraphics[width = 0.30\linewidth]{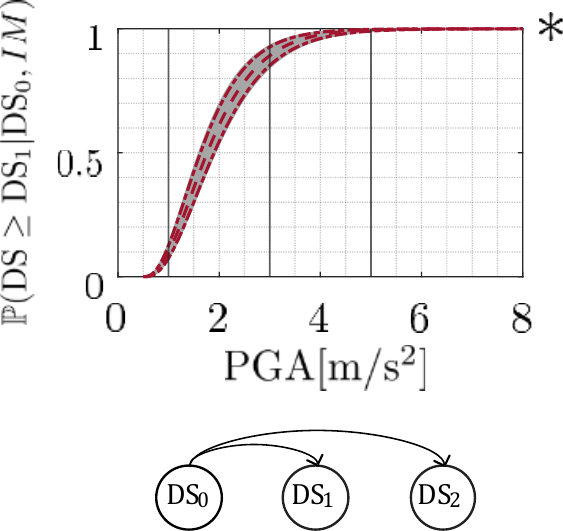}
	\caption{The commonly used ``fragility function'' evaluated as the exceedance probability, i.e., $\mathbb{P}(\mathrm{DS}\ge\mathrm{DS_1}|\mathrm{DS_0},IM)$, for PGA as IM.}
	\label{fig:ch3_state_01_PGA}
\end{figure}

\section{- SPIF \#2 support material.} \label{app:D_SPIF_collection}
\setcounter{figure}{0}                       
\renewcommand\thefigure{C.\arabic{figure}}   
\subsection{Collection of the optimal state-dependent functions for each transition state.}
\begin{figure}[!ht]
	\centering
	\includegraphics[scale = 0.55]{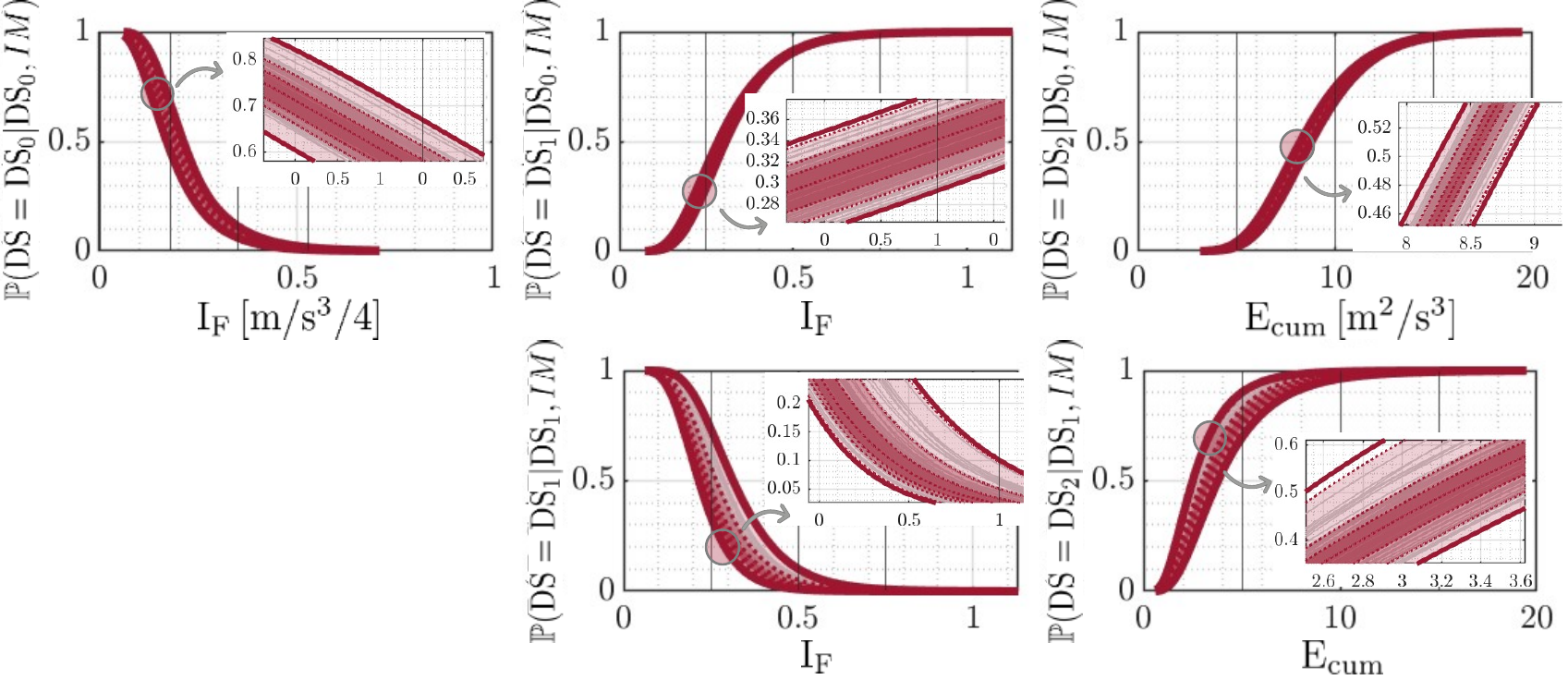}
	\caption{Bootstrap-PCE state-dependent fragility curves of the SPIF \#2 vertical tank: dark-red thick lines stand for the 50\%, 90\% and 99\% confidence bound, along with lighter to darker shaded areas.}
	\label{fig:appC_PCE_fragility}
\end{figure}

\noindent The optimal IMs for each transition state were determined according to the $\mathrm{\beta_{eff}}$ index of Eq.~\ref{eq:betaIndex}. Table~\ref{tab:ch4_spif_beta} collects the top-five-ranked IMs for each transition state.
It is possible to notice that the Fajfar intensity $\mathrm{I_{F}}$, the equipment relative average spectral acceleration $\operatorname{\mathrm{E-ASA_{R67}}}$, and the cumulated energy $\mathrm{E_{cum}}$ are the optimal IMs that occur more often among the transition states. Specifically, the $\mathrm{E_{cum}}$ and the $\operatorname{\mathrm{E-ASA_{R67}}}$ are directly correlated with acceleration and energy content (see how they are defined in Table~\ref{tab:ch3_feature_IM}); whilst the $\mathrm{I_{F}}$ is correlated with velocity terms. Besides, the $\operatorname{\mathrm{E-ASA_{R67}}}$ is the only one that is repeated on all the transition states. As deeply investigated in \cite{bib:DeBiasio15}, this IM is particularly suited to capture the drop and frequency shifts of equipment characterised by a significant spectral acceleration close to the dominant frequency interval of the main structure.  
These observations agree with the experimental evidence described in~\cite{bib:nardin2022}. Indeed, the shake table data revealed a significant positive correlation between the maximum floor spectral acceleration $S_{a,floor}(T_1)$ and the $\operatorname{\mathrm{E-ASA_{R67}}}$ for the vertical tank.
Thus, Figure~\ref{fig:appC_PCE_fragility} reports the state-dependent fragility functions for the industrial component with the optimal IM for each transition state. Dark-red lines and the associated shaded areas highlight the 50\%, 90\% and 99\% confidence bounds, respectively. As the structure reaches the damage level $\mathrm{DS_2}$, $\mathrm{E_{cum}}$ is the optimal IM for both pristine and damaged initial level conditions.  It can be noted that given a damage limit state to attain, i.e., (graphically) for elements of the same column, the magnitude of IM required to reach the same exceedance probability is lower with the damaged initial conditions. For instance, the 50\% probability of exceedance $\mathrm{DS_2}$ given $\mathrm{DS_0}$ and $\mathrm{DS_1}$ is associated to $\mathrm{E_{cum}} = 9$ and $\mathrm{E_{cum}} = 3$, respectively. 
\begin{table*}[htbp]
	\centering
	\caption{Top five-ranked IM $\beta_{\mathrm{eff}}$ efficiency indices for each transition state.}  \label{tab:ch4_spif_beta}
	\scalebox{0.65}{
		\begin{tabular}{lrrlrrlrrlrrlr}
			\cmidrule{1-2}\cmidrule{4-5}\cmidrule{7-8}\cmidrule{10-11}\cmidrule{13-14}    \multicolumn{1}{c}{\textbf{Frag. \apo0-0\apo}} & \multicolumn{1}{c}{\boldmath{}\textbf{$\mathrm{\beta_{eff}}$}\unboldmath{}} &       & \multicolumn{1}{c}{\textbf{Frag. \apo0-1\apo}} & \multicolumn{1}{c}{\boldmath{}\textbf{$\mathrm{\beta_{eff}}$}\unboldmath{}} &       & \multicolumn{1}{c}{\textbf{Frag. \apo0-2\apo}} & \multicolumn{1}{c}{\boldmath{}\textbf{$\mathrm{\beta_{eff}}$}\unboldmath{}} &       & \multicolumn{1}{c}{\textbf{Frag. \apo1-1\apo}} & \multicolumn{1}{c}{\boldmath{}\textbf{$\mathrm{\beta_{eff}}$}\unboldmath{}} &       & \multicolumn{1}{c}{\textbf{Frag. \apo1-2\apo}} & \multicolumn{1}{c}{\boldmath{}\textbf{$\mathrm{\beta_{eff}}$}\unboldmath{}} \\
			\multicolumn{1}{c}{\textbf{Optimal IM}} & \multicolumn{1}{c}{\textbf{index}} &       & \multicolumn{1}{c}{\textbf{Optimal IM}} & \multicolumn{1}{c}{\textbf{index}} &       & \multicolumn{1}{c}{\textbf{Optimal IM}} & \multicolumn{1}{c}{\textbf{index}} &       & \multicolumn{1}{c}{\textbf{Optimal IM}} & \multicolumn{1}{c}{\textbf{index}} &       & \multicolumn{1}{c}{\textbf{Optimal IM}} & \multicolumn{1}{c}{\textbf{index}} \\
			\cmidrule{1-2}\cmidrule{4-5}\cmidrule{7-8}\cmidrule{10-11}\cmidrule{13-14}    $\mathrm{I_F}$ & 5.00E-04 &       & $\mathrm{I_F}$ & 2.00E-04 &       & $\mathrm{E_{cum}}$ & 1.02E-04 &       & $\mathrm{I_F}$ & 3.00E-04 &       & $\mathrm{E_{cum}}$ & 6.96E-04 \\
			$\mathrm{PGA}$ & 1.24E-02 &       & $\mathrm{E_{cum}}$ & 7.00E-04 &       & $\mathrm{I_A}$ & 9.25E-06 &       & $\mathrm{PGA}$ & 6.90E-03 &       & $\mathrm{Sv(T_1)}$ & 4.46E-03 \\
			$\mathrm{I_C}$ & 3.75E-02 &       & $\mathrm{I_A}$ & 3.80E-03 &       & $\mathrm{I_C}$ & 6.58E-06 &       & $\mathrm{E_{cum}}$ & 4.14E-02 &       & $\mathrm{PGA}$ & 3.15E-03 \\
			$\mathrm{I_{RG,a}}$ & 5.35E-02 &       & $\operatorname{\mathrm{E-ASA_{R67}}}$ & 6.70E-03 &       & $\mathrm{I_{CM}}$ & 5.82E-06 &       & $\mathrm{I_{RG,a}}$ & 7.94E-02 &       & $\mathrm{E-ASA_{R67}}$ & 2.58E-03 \\
			$\mathrm{E-ASA_{R67}}$ & 8.51E-02 &       & $\mathrm{PGA}$ & 1.60E-02 &       & $\operatorname{\mathrm{E-ASA_{R67}}}$ & 1.81E-06 &       & $\operatorname{\mathrm{E-ASA_{R67}}}$ & 9.55E-02 &       & $\operatorname{\mathrm{E-ASA_{R100}}}$ & 6.19E-02 \\
			\cmidrule{1-2}\cmidrule{4-5}\cmidrule{7-8}\cmidrule{10-11}\cmidrule{13-14}    
		\end{tabular}%
	}
\end{table*}%
\FloatBarrier
\subsection{State-dependent fragility: $\mathbb{P}(\mathrm{DS}\geq \mathrm{DS_1}|\mathrm{DS_0}, IM)$.} \label{app:D_SPIF_01_PGA}
\begin{figure}[!ht]
	\centering
	\includegraphics[width=0.30\linewidth]{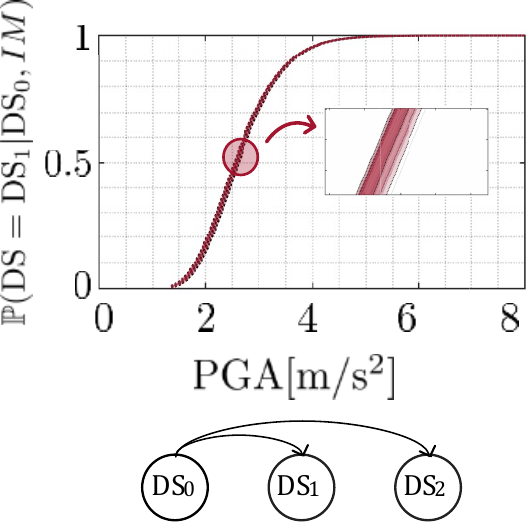}
	\caption{The commonly used ``fragility function'' evaluated as the exceedance probability, i.e., $\mathbb{P}(\mathrm{DS}\ge\mathrm{DS_1}|\mathrm{DS_0},IM)$, with PGA as IM.}
	\label{fig:D_SPIF_01_PGA}
\end{figure}

\newpage
\bibliography{bibliography.bib}

\end{document}

%% file: Tables/app_A_IM_distribution.tex
\begin{table*}[h!]
  \centering
    \caption{Histograms and distributional models inferred for each IM.}{
    \label{tab:app_A_IM_distribution}%
     \scalebox{0.8}{
        \renewcommand{\arraystretch}{1.1} 
        \begin{tabular}{clllc}
    \toprule
    \textbf{Index } & \textbf{ Name } & \textbf{ Type        } & \textbf{  Parameters             } & \multicolumn{1}{l}{\textbf{ Moments              }} \\
    \midrule
          &       &       &       &  \\
    1     & $\mathrm{PGA}$ &  Gumbel       &   2.027e+00, 6.236e-01    &  2.387e+00, 7.998e-01 \\
    2     & $\mathrm{PGV}$ &  Gumbel       &   1.408e-01, 5.683e-02    &  1.736e-01, 7.289e-02 \\
    3     & $\mathrm{PGD}$ &  Gamma        &   3.853e+01, 3.489e+00    &  9.055e-02, 4.848e-02 \\
    4     & $\mathrm{Sd(T_1)}$ &  Gumbel       &   1.447e-02, 6.679e-03    &  1.833e-02, 8.566e-03 \\
    5     & $\mathrm{Sd(T_2)}$ &  Gumbel       &   1.031e-02, 4.263e-03    &  1.277e-02, 5.468e-03 \\
    6     & $\mathrm{Sd(T_3)}$ &  Gumbel       &   7.792e-03, 2.635e-03    &  9.313e-03, 3.379e-03 \\
    7     & $\mathrm{Sd(T_4)}$ &  LogNormal    &   -5.551e+00, 3.517e-01   &  4.133e-03, 1.500e-03 \\
    8     & $\mathrm{Sv(T_5)}$ &  LogNormal    &   -6.831e+00, 3.239e-01   &  1.137e-03, 3.782e-04 \\
    9     & $\mathrm{Sv(T_1)}$ &  Gumbel       &   1.814e-01, 8.370e-02    &  2.297e-01, 1.074e-01 \\
    10    & $\mathrm{Sv(T_2)}$ &  Gumbel       &   1.835e-01, 7.591e-02    &  2.274e-01, 9.736e-02 \\
    11    & $\mathrm{Sv(T_3)}$ &  Gumbel       &   1.958e-01, 6.622e-02    &  2.341e-01, 8.493e-02 \\
    12    & $\mathrm{Sv(T_4)}$ &  LogNormal    &   -1.815e+00, 3.517e-01   &  1.732e-01, 6.285e-02 \\
    13    & $\mathrm{Sv(T_5)}$ &  LogNormal    &   -2.690e+00, 3.239e-01   &  7.156e-02, 2.380e-02 \\
    14    & $\mathrm{Sa(T_1)}$ &  Gumbel       &   2.273e+00, 1.049e+00    &  2.878e+00, 1.345e+00 \\
    15    & $\mathrm{Sa(T_2)}$ &  Gumbel       &   3.268e+00, 1.352e+00    &  4.048e+00, 1.733e+00 \\
    16    & $\mathrm{Sa(T_3)}$ &  Gumbel       &   4.922e+00, 1.664e+00    &  5.883e+00, 2.134e+00 \\
    17    & $\mathrm{Sa(T_4)}$ &  LogNormal    &   1.920e+00, 3.517e-01    &  7.260e+00, 2.634e+00 \\
    18    & $\mathrm{Sa(T_5)}$ &  LogNormal    &   1.452e+00, 3.239e-01    &  4.502e+00, 1.497e+00 \\
    19    & $\mathrm{\mathrm{I_A}}$ &  LogNormal    &   -6.729e-01, 5.712e-01   &  6.006e-01, 3.730e-01 \\
    20    & $\mathrm{E_{cum}}$ &  LogNormal    &   1.342e+00, 5.453e-01    &  4.439e+00, 2.612e+00 \\
    21    & $\mathrm{I_{RG,a}}$ &  Gumbel       &   4.151e+00, 1.222e+00    &  4.856e+00, 1.567e+00 \\
    22    & $\mathrm{I_{RG,v}}$ &  LogNormal    &   -5.098e-01, 2.745e-01   &  6.237e-01, 1.745e-01 \\
    23    & $\mathrm{T_d}$ &  Gaussian     &   8.696e+00, 1.637e+00    &  8.696e+00, 1.637e+00 \\
    24    & $\mathrm{RMS(\ddot{u}(t))}$ &  LogNormal    &   -7.590e-01, 2.845e-01   &  4.875e-01, 1.415e-01 \\
    25    & $\mathrm{IC}$ &  LogNormal    &   -6.627e-02, 4.070e-01   &  1.017e+00, 4.315e-01 \\
    26    & $\mathrm{CAV}$ &  Gamma        &   9.262e-03, 9.638e+00    &  1.041e+03, 3.352e+02 \\
    27    & $\mathrm{I_{CM}}$ &  LogNormal    &   2.178e+00, 4.817e-01    &  9.915e+00, 5.067e+00 \\
    28    & $\mathrm{ASA_{40}}$ &  Gamma        &   1.122e-01, 8.580e+00    &  7.647e+01, 2.611e+01 \\
    29    & $\mathrm{ASI}$ &  LogNormal    &   6.025e+00, 2.910e-01    &  4.315e+02, 1.283e+02 \\
    30    & $\mathrm{EPA}$ &  Gamma        &   4.737e-02, 1.134e+01    &  2.393e+02, 7.108e+01 \\
    31    & $\mathrm{I_{v/a}}$ &  Gamma        &   1.343e+02, 9.941e+00    &  7.400e-02, 2.347e-02 \\
    32    & $\mathrm{I_{F}}$ &  Gumbel       &   2.407e-01, 9.524e-02    &  2.957e-01, 1.221e-01 \\
    33    & $\mathrm{F_m}$ &  Gaussian     &   4.649e+00, 4.801e-01    &  4.649e+00, 4.801e-01 \\
    34    & $\mathrm{\dot{F}_m}$ &  GumbelMin    &   -2.667e-04, 2.347e-04   &  -4.022e-04, 3.011e-04 \\
    35    & $\mathrm{FAS_{area}}$ &  Exponential  & 3362  &  2.974e-01, 2.974e-01 \\
    36    & $\mathrm{E-ASA_{R_{40}}}$ &  LogNormal    &   2.371e+00, 3.504e-01    &  1.138e+01, 4.115e+00 \\
    37    & $\mathrm{E-ASA_{R_{67}}}$ &  LogNormal    &   3.283e+00, 3.387e-01    &  2.823e+01, 9.842e+00 \\
    38    & $\mathrm{E-ASA_{R_{80}}}$ &  LogNormal    &   3.561e+00, 3.325e-01    &  3.719e+01, 1.271e+01 \\
    39    & $\mathrm{E-ASA_{R_{100}}}$ &  LogNormal    &   3.917e+00, 3.229e-01    &  5.296e+01, 1.756e+01 \\
    40    & $\mathrm{E-ASA_{R_{150}}}$ &  LogNormal    &   4.495e+00, 3.110e-01    &  9.398e+01, 2.995e+01 \\
    41    & $\mathrm{E-ASA_{R_{200}}}$ &  LogNormal    &   4.879e+00, 3.056e-01    &  1.378e+02, 4.313e+01 \\
    \bottomrule
    \end{tabular}%
    }}
\end{table*}%